\documentclass[journal]{IEEEtran}

\usepackage[dvipsnames]{xcolor}
\usepackage{graphicx}
\usepackage{epstopdf}
\usepackage{float}
\usepackage{color}
\usepackage{amsmath}
\usepackage{array}
\usepackage{relsize}
\usepackage{amsmath,amssymb,latexsym}
\usepackage{stfloats}
\usepackage{caption}
\usepackage{subcaption}
\usepackage{tikz}
\usepackage{tabularx}
\usepackage{multirow}
\usepackage{harveyballs}
\usepackage[font=small]{caption}

\begin{document}

\title{Energy-Efficient Transmission Range and Duration for Cognitive Radio Sensor Networks}

\author{Mustafa~Ozger,~
		Ecehan~B.~Pehlivanoglu,~
		and~Ozgur~B.~Akan,~\IEEEmembership{Fellow,~IEEE}
\thanks{M. Ozger is with the School of Electrical Engineering and Computer Science, KTH Royal Institute of Technology, Stockholm, Sweden. e-mail: ozger@kth.se.}
\thanks{E. B. Pehlivanoglu and O. B. Akan are with Next-generation and Wireless Communications Laboratory (NWCL), the Department of Electrical and Electronics Engineering, Koc University, Istanbul, 34450, Turkey. e-mail: epehlivanoglu@ku.edu.tr.}
\thanks{O. B. Akan is also with Internet of Everything (IoE) Group, Electrical Engineering Division, Department of Engineering, University of Cambridge, CB3 0FA Cambridge, UK. e-mail: oba21@cam.ac.uk.}
}
\maketitle

\begin{abstract}
Cognitive Radio (CR) promises an efficient utilization of radio spectrum resources by enabling dynamic spectrum access to overcome the spectrum scarcity problem. Cognitive Radio Sensor Networks (CRSNs) are one type of Wireless Sensor Networks (WSNs) equipped with CR capabilities. CRSN nodes need to operate energy-efficiently to extend network lifetime due to their limited battery capacity. {In this paper, for the first time in literature, we formulate the problem of finding a common energy-efficient transmission range and transmission duration for all CRSN nodes and network deployment that would minimize the energy consumed per goodput per meter toward the sink in a greedy forwarding scenario. Results reveal non-trivial relations for energy-efficient CRSN transmission range and duration as a function of nine critical network parameters such as primary user activity levels. These relations provide valuable insights for detailed CRSN designs prior to deployment.}
\end{abstract}


\IEEEpeerreviewmaketitle

\section{Introduction}

The proliferation of wireless devices has caused spectrum scarcity since the available spectrum is not efficiently utilized with the current fixed spectrum allocation approach. Dynamic Spectrum Access (DSA) is a paradigm to overcome such inefficiencies, by allowing unlicensed users to leverage the idle licensed spectrum. Cognitive Radio (CR) has been proposed as the key enabling technology for the realization of DSA. CR is applied to Wireless Sensor Networks (WSNs) to form Cognitive Radio Sensor Networks (CRSNs), making the most use of available spectrum for sensor networks without the burden of spectrum licensing costs \cite{crsn.survey.akan}.

Given the energy constraints of unlicensed CRSN nodes, called Secondary Users (SUs), communication between a CRSN source node and the remote sink takes place through multiple hops under DSA. Successful frame transmission under DSA in each hop depends on certain requirements. Firstly, SU transmitter and receiver nodes need to be in the transmission range of each other (i.e., they need to be topologically connected one-hop neighbors), and both need to correctly identify the same portion of the spectrum as idle through spectrum sensing. Secondly, a licensed user, called Primary User (PU), should not arrive on that portion of the spectrum during the SU transmission. In case of a PU arrival, the SU pair need to restart sensing the spectrum in the next time slot and attempt retransmission if they can seize some portion of the spectrum. In this context, transmission range and duration have a non-trivial impact on the energy-efficient operations of a CRSN. A larger transmission range results in a fewer number of hops toward the remote sink yet faces higher energy consumption at each hop and a bigger potential number of PUs that may interrupt the SU communications. On the other hand, a smaller transmission range decreases energy consumption at each hop and reduces the number of interfering PUs but increases the number of hops to reach the remote sink, hence might potentially increase overall source-to-sink energy consumption. Additionally, a longer transmission duration can deliver more bits in one transmission slot for a given bit rate, yet is more susceptible to incoming PUs on the accessed part of the spectrum during transmission.

{Transmission range is often a set parameter before sensor network deployment as simple sensor nodes may not have complex power control capabilities \cite{fixed.range.gandham, fixed.range.bettstetter}. Furthermore, power control may not be cost-effective under DSA  due to limited energy-saving potential and added protocol complexity \cite{fixed.range.de}.} Energy-efficient transmission range is even more important in the CRSN context, given its impact on successful frame transmission under DSA schemes {limited battery capacity}. {Existing works that shed light on transmission range for ad hoc wireless networks and WSNs prior to their deployment \cite{fixed.range.deng, fixed.range.chen} are not directly applicable to the CRSN context given the intermittent nature of communication under DSA. Furthermore, there is no study dedicated to finding transmission range and duration for all nodes in CRSN as design parameters. Additionally, while it could be possible for CRSN to tune its transmission duration, finding an energy-efficient duration for different scenarios prior to deployment is key. In fact, \cite{crsn.oto} considers this from the perspective of energy-efficient packet size. To that end, energy-efficient transmission range and duration can be studied with a composite metric on energy consumed per goodput per meter progress toward the sink at each hop.}

Recent research on CRSNs focused mostly on spectrum sensing \cite{spectrum.sensing.zhang, spectrum.sensing.maryam, spectrum.sensing.new}, spectrum decision and handoff \cite{crsn.channel.management.han, hand_off}, clustering \cite{ozger},  physical layer studies on adaptive modulation \cite{physical.gao}, CR-enabled Internet of Things \cite{cr-iot} and channel bonding for CRSN \cite{physical.rehmani} and attempts on MAC layer \cite{mac.crsn}, network layer \cite{network.ge} and transport layer \cite{tcp_crsn} protocols and solutions. {In addition, full-duplex communication with non-orthogonal multiple access is also investigated for security and reliability of in CR-enabled networks \cite{fullduplex1, fullduplex2}. Also, the authors in \cite{EH_resource} focus on maximization of minimum data rate in IoT networks with energy harvesting and CR capabilities. Although these studies are important to realize CRSNs, there is clearly a need to study homogeneous energy-efficient transmission range and duration for CRSN nodes. Since the sensors are simple devices with limited computational and battery capacities, changing transmission range and duration under the dynamic spectrum access environment would be prohibitive. Hence, our analysis serves as valuable guidelines for designers on how to set a fixed transmission range and duration prior to the deployment.}

In this paper, for the first time in literature, we formulate the problem of finding energy-efficient homogeneous transmission range and duration for the nodes in CRSN deployments. To that end, we define two clear objectives, which then translate into the network model and consequent analyses to shed light on energy-efficient transmission range and duration for CRSNs:

\begin{itemize}
\item \textbf{\emph{Objective 1: Spectrum-aware transmissions}}: Transmission range and duration should enable efficient transmission in each hop. The success of frame transmission on each hop of a CRSN depends on (i) Presence of a spectrum opportunity and its correct sensing (after false alarms and mis-detections are accounted for) both by the transmitter and the receiver; (ii) No PU arrival during the entire transmission duration on the accessed spectrum portion; (iii) Correct reception of all bits on the receiver side under the unreliable radio link model. 
    
\item \textbf{\emph{Objective 2: Energy-efficient operations}}: Transmission range and duration should ensure overall energy-efficiency of the CRSN. This depends on the minimization of energy consumed per goodput per meter progress toward sink at each hop, hence increasing overall network energy-efficiency.
      
\end{itemize}

{Contributions of our paper are as follows. Following these two core objectives, we first clearly lay our non-binding assumptions on the system model. Secondly, we characterize the transmissions in CRSN under DSA; deriving the probability of successful frame transmission at each attempt between one-hop neighbor CRSN nodes, extending the study in \cite{spectrum.opportunity.ozger}. Thirdly, we derive expected hop progress from the source toward the sink at each hop (i.e. meters taken toward the sink at each hop). Given a homogeneous transmission range, we lay out the expected hop progress (i.e. the expected progress toward the sink at each hop) as well as the expected hop distance (i.e. the expected distance between one-hop neighbors) in a greedy forwarding scheme. From the expected hop distance, we reveal the expected energy consumption until successful frame transmission under different spectrum utilization scenario probabilities. Finally, we derive the energy consumed per goodput per-hop progress in meter toward the sink and reveal optimal transmission range and transmission duration values thereof.} The main contribution of this paper is to reveal energy-efficient homogeneous transmission range and transmission duration for CRSN deployment and to investigate its relationship with the following critical network parameters:
path loss exponent ($\kappa$),
 PU death and birth rates ($\alpha$ and $\beta$),
 PU node density ($\rho_p$),
 Radius of PU guardring protection zone ($r_p$),
 Reference Signal-to-Noise Ratio (SNR) at SU receivers for correct demodulation ($\gamma_0$),
 Power consumed by SU during spectrum sensing ($P_s$),
 Maximum SU-PU collision probability that a PU can tolerate ($P_{col}$),
 Ratio of PU signal variance ($\sigma_p^2$) to noise variance ($\sigma_n^2$).

The remainder of the paper is organized as follows. In Section \ref{SystemModel}, the system model to investigate the energy-efficient transmission range and duration for CRSN is introduced. Derivations related to DSA characteristics, energy consumption dynamics of each node, successful frame transmission probability under the DSA scheme, and energy consumption thereof are presented in Section \ref{AnalysisFramework}. For a given homogeneous transmission range, we study the expected hop progress toward the sink as well as expected hop distance in a greedy forwarding scheme in Section \ref{ProblemFormulation}. Using this result, we derive the non-closed form equation for energy-efficient homogeneous transmission range and duration for CRSN in the same section. The numerical results are presented and discussed in Section \ref{NumericalResults}, and the paper is concluded in Section \ref{Conclusions}.

\section{System Model}
\label{SystemModel}

In our system model, there are two types of network devices. The PUs have exclusive access to the licensed spectrum. On the other hand, the SUs access these licensed bands opportunistically, i.e., whenever there are no PU transmissions on them. The following features are incorporated in our analysis:

\begin{itemize}

\item PUs coexist with SUs on a 2-dimensional circular area (called $\mathcal{F}$, of radius $\Gamma$). 

\item Both PUs and SUs are randomly deployed over $\mathcal{F}$ with 2-D homogeneous Poisson Point Processes (PPP) with means $\rho_p$ and $\rho_s$, respectively.

\item CRSN has a flat structure. Each SU node senses its environment and transmits event information to the sink in multiple hops. The sink resides at the center of $\mathcal{F}$.

\item Simple greedy forwarding is assumed as routing mechanism \cite{greedy.forwarding.karp}. Hence, an SU's next hop neighbor to transmit its field sensing information is assumed to be the neighbor within a range of $r_s$ that is closest to the remote sink.

\item SU nodes are assumed to know their neighbors' locations with respect to the remote sink, through a geographical location service \cite{gls.li}.

\item SUs have a common predetermined channel list, which they follow during periodic spectrum sensing operations \cite{spectrum.opportunity.bicen}. This list may be provided by a centralized spectrum sharing mechanism. Formation of the optimal channel list is beyond the scope of this paper, and its effect on our analysis is neutralized assuming all channels have the same PU density, PU activity model and bandwidth.

\item Time is slotted for SU-SU communication. In each slot, two one-hop neighbors that are willing to communicate sense a licensed channel from their channel list, for Spectrum Access (SA). If they run out of channels in their list, they restart in a circular fashion.

\item Two SU neighbors have a Spectrum Opportunity (SO) if i) they are not within the guardring of any PU at all times OR ii) the PUs in whose guardring they reside are all inactive during spectrum sensing. The guardring of a PU is centered at that PU's location, with radius $r_p$.

\item Spectrum sensing might be erroneous. In case both SUs correctly sense the SO, that event is called Successful Spectrum Access (SSA).

\item In case any of the two SU neighbors sense (either correctly or erroneously) a PU presence during spectrum sensing on the licensed channel, they perform spectrum handoff, moving to the next channel in their channel list, in a circular fashion. 

\item When two SU neighbors carry out SA (whether successful or not), \textit{Frame Transmission (FT)} is started. In case no PU arrives during frame transmission, Successful Frame Transmission (SFT) takes place. Otherwise, the SU packet is not correctly delivered, and both SUs perform spectrum handoff to the next channel at the end of the transmission. We assume there is no dedicated radio to notify the SUs that a PU has arrived on the channel they communicate \cite{spectrum.opportunity.bicen}, hence in such a case SUs perform spectrum handoff at the of the transmission slot $\tau_f$.

\item The PU activity is modeled as exponentially distributed inter-arrivals, with a two state birth-death process with death rate $\alpha$ and birth rate $\beta$. As a result of this birth-death process, the licensed spectrum portion under scrutiny can have ON and OFF states. An ON (Busy) state represents the period where this spectrum portion is occupied by PUs, whereas an OFF (Idle) state represents the period where it is unused. The length of the ON and OFF periods are exponentially distributed \cite{crsn.oto, optimal.spectrum.sensing.lee}.

\end{itemize}

\section{Analysis Framework}
\label{AnalysisFramework}

\subsection{Spectrum Sensing Events and Their Probabilities}
\label{DynamicSpectrumAccess}
\begin{figure}[!ht]
	\centering
	\includegraphics[width=0.7\columnwidth]{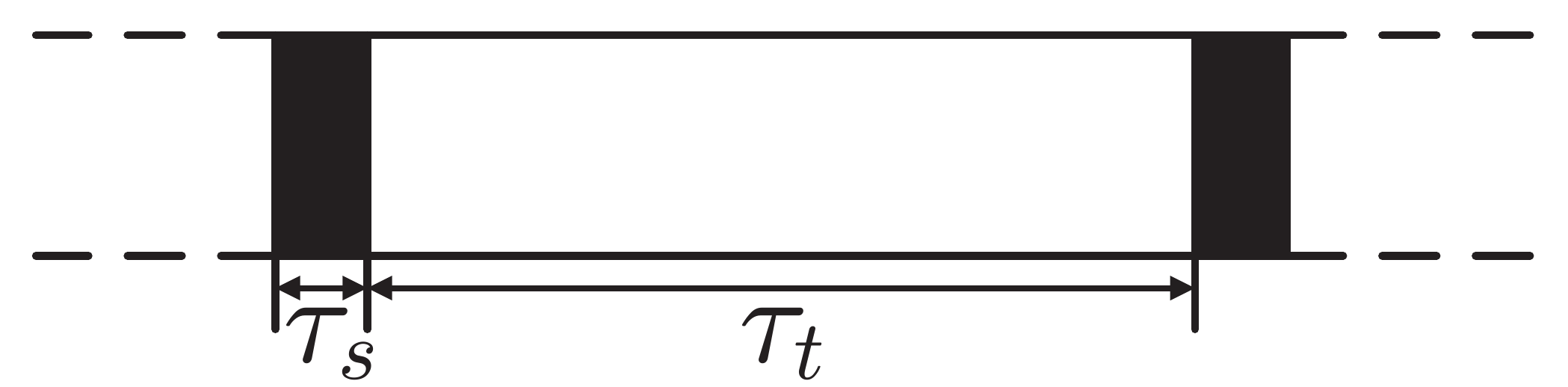}
	\caption{ The time slot structure.}
	\label{timeslot}
\end{figure}

The time is slotted, with two periods in each slot. The first period is reserved for spectrum sensing ($\tau_s$), and the subsequent period is for the frame transmission ($\tau_t$).  Total duration of one slot is $\tau_f = \tau_s + \tau_t$, as depicted in Fig. \ref{timeslot}.

$T_{idle}$ and $T_{busy}$ are the random variables that define the idle and busy time periods of each licensed channel with probability distribution functions $f_{T_{idle}}(t)=\beta e^{-t \beta}$ and $f_{T_{busy}}(t)=\alpha e^{-t \alpha}$, as per PU activity model explained in Section \ref{SystemModel} \cite{optimal.spectrum.sensing.lee}.
Resultant channel probabilities are $P_{idle}=\alpha/(\alpha+\beta)$ and $P_{busy}=\beta/(\alpha+\beta)$, respectively. 

Due to its practicality, energy detection based spectrum sensing is assumed, with non-zero false alarm and mis-detection probabilities. Let the hypotheses $\mathcal{H}_0$ and $\mathcal{H}_1$ define idle and busy channel conditions, respectively. Hence, the probability of false alarm is
\begin{equation}
P_{fa}= Pr[Busy, \mathcal{H}_0]= P_{idle} \mathcal{Q}\Big(\frac{\delta-2\tau_s B \sigma_n^2}{\sqrt{4 \tau_s B \sigma_n^4}}\Big),
\label{eq:P_fa}
\end{equation}
where $B$ is the bandwidth of the channel, $\delta$ is the detection threshold, $\sigma_n^2$ is the noise variance, and $\mathcal{Q}(.)$ is Q-function \cite{optimal.spectrum.sensing.lee}. The probability of correct busy detection is given as
\begin{equation}
P_{d}= Pr[Busy, \mathcal{H}_1]= P_{busy} \mathcal{Q}\Big(\frac{\delta-2\tau_s B (\sigma_p^2+\sigma_n^2)}{\sqrt{4 \tau_s B (\sigma_p^2+\sigma_n^2)^2}}\Big),
\label{eq:P_d}
\end{equation}
where $\sigma_p^2$ is the PU signal variance \cite{optimal.spectrum.sensing.lee}. Probabilities of correct detection of idle spectrum and mis-detection are given as follows, respectively: 
\begin{equation}
P_v=P_{idle}- P_{fa},
\label{eq:P_v}
\end{equation}
\begin{equation}
P_{md}=P_{busy}- P_{d}.
\label{eq:P_md}
\end{equation}

\subsection{Energy Consumption Dynamics}
\label{EnergyConsumptionDynamics}

Energy consumed in each hop has three components. The first component is $E_s$, consumed for spectrum sensing during $\tau_s$ at the beginning of each slot. Both the transmitter SU and the receiver SU in a one-hop transmission consume $E_s=P_s \tau_s$ amount of energy for each spectrum sensing operation.

The second component is $E_t$, energy consumed for transmission which {is a function of $r_s$ and $\tau_t$, and} can be formulated as \cite{fixed.range.chen, energy.model.zhang}
\begin{equation}
E_t=(q_1 r_{s}^{\kappa} + q_2)\tau_t,
\label{eq:E_t}
\end{equation} where $\kappa$ is the path loss exponent, $r_s$ is the transmission range and $\tau_t$ is the transmission period in a single time slot. $q_1$ and $q_2$ can be defined as \cite{fixed.range.chen}
\begin{equation}
q_1=\displaystyle\frac{\gamma_0 N_{rx} N_0 B (\frac{4 \pi}{\lambda})^{\kappa} 10^{\kappa}}{G_{a} \eta_{amp}},
\end{equation}
\begin{equation}
q_2=\displaystyle P_{elec},
\end{equation}
where $\gamma_0$ is the minimum required signal-to-noise (SNR) ratio at the receiver's demodulator, $N_{rx}$ is the noise figure of the receiver, $N_0$ is the thermal noise in a 1 Hz bandwidth, $B$ is the channel bandwidth, $\lambda$ is the signal wavelength, $\eta_{amp}$ is the transmitter amplifier efficiency, $G_{a}$ is the antenna gain and $P_{elec}$ is the power consumed by the device. 

The third component is $E_r$, energy consumed on receiver side, formulated as $E_r=P_{rx} \tau_t$, where  $P_{rx}$ is the receiver power consumption.

\begin{table*}[!t]
\scriptsize

  \centering
  \caption{{Scenarios for utilization of spectrum opportunities in the system model, probabilities and energy consumption thereof.}}
\bgroup
\def\arraystretch{1.8}
\resizebox{0.99\textwidth}{!}{%
\begin{tabular}{cl|l|l|l|l||l|l|l|l|l|l|}
\cline{3-10}
& & \multicolumn{4}{ |c|| }{\textbf{Spectrum Access (SA)}}  & \multicolumn{4}{ c|| }{\textbf{Frame Transmission (FT)}} \\
\cline{3-10}

& & \multicolumn{2}{ |c| }{\textbf{TX sensing}} & \multicolumn{2}{ c|| }{\textbf{RX sensing}} & \multicolumn{2}{ c| }{\textbf{PU arrival in $\tau_t$}} & \multicolumn{2}{ c|| }{\textbf{Reliability in $\tau_t$}} \\ \cline{1-12} 

\multicolumn{1}{ |m{0.4 cm}||  }{\textbf{S \#}} & \multicolumn{1}{ m{0.6 cm}|| }{\textbf{Init. state}} & \multicolumn{1}{ |c|  }{\textit{Event}} & \multicolumn{1}{ |c| }{\textit{Prob.}} & \multicolumn{1}{ |c| }{\textit{Event}} & \multicolumn{1}{ |c||  }{\textit{Prob.}} & \multicolumn{1}{ c|  }{\textit{Event}} & \multicolumn{1}{ c|  }{\textit{Prob.}} & \multicolumn{1}{ c|  }{\textit{Event}} & \multicolumn{1}{ c||  }{\textit{Prob.}} & \multicolumn{1}{ c|  }{\textbf{Total probability}} & \multicolumn{1}{ c|  }{\textbf{Cons. energy}}\\ \cline{1-12}

\multicolumn{1}{ |c|| }{S1} & \multicolumn{1}{ |l|| }{$P_{idle}$} & FA$^1$ & $\overline{P_{fa}}$ & FA$^1$ & $\overline{P_{fa}}$ & - & 1 & - & \multicolumn{1}{ |l|| }{1} & $P_{idle} \overline{P_{fa}}^2$ & $2E_{s}$ \\ \cline{1-12}

\multicolumn{1}{ |c|| }{S2} & \multicolumn{1}{ |l|| }{$P_{idle}$} & FA$^1$ & $\overline{P_{fa}}$ & CVD$^2$ & $\overline{P_{v}}$ & - & 1 & - & \multicolumn{1}{ |l|| }{1} & $P_{idle} \overline{P_{fa}}$ $\overline{P_{v}}$ & $2E_{s}$ \\ \cline{1-12}

\multicolumn{1}{ |c|| }{S3} & \multicolumn{1}{ |l|| }{$P_{idle}$} & CVD$^2$ & $\overline{P_{v}}$ & FA$^1$ & $\overline{P_{fa}}$ & - & 1 & - & \multicolumn{1}{ |l|| }{1} & $P_{idle} \overline{P_{fa}}$ $\overline{P_{v}}$ & $2E_{s}$ \\ \cline{1-12}

\multicolumn{1}{|c|| }{S4} & \multicolumn{1}{ |l|| }{$P_{idle}$} & CVD$^2$ & $\overline{P_{v}}$ & CVD$^2$ & $\overline{P_{v}}$ & $\geq 1$ PU$^5$ & $1-P_{np}$ & - & \multicolumn{1}{ |l|| }{1} & $P_{idle} \overline{P_{v}}^2 (1-P_{np})$ & $2E_{s}+E_t+E_r$ \\ \cline{1-12}

\multicolumn{1}{ |c|| }{S5} & \multicolumn{1}{ |l|| }{$P_{idle}$} & CVD$^2$ & $\overline{P_{v}}$ & CVD$^2$ & $\overline{P_{v}}$ & No PU & $P_{np}$ & $\geq 1$, error & \multicolumn{1}{ |l|| }{$1-P_{r}$} & $P_{idle} \overline{P_{v}}^2 P_{np}(1-P_{r})$ & $2E_{s}+E_t+E_r$ \\ \cline{1-12}

\multicolumn{1}{ |c|| }{S6} & \multicolumn{1}{ |l|| }{$P_{idle}$} & CVD$^2$ & $\overline{P_{v}}$ & CVD$^2$ & $\overline{P_{v}}$ & No PU & $P_{np}$ & No error & \multicolumn{1}{ |l|| }{$P_{r}$} & $P_{idle} \overline{P_{v}}^2 P_{np}P_{r}$ & $2E_{s}+E_t+E_r$ \\ \cline{1-12}

\multicolumn{1}{ |c|| }{S7} & \multicolumn{1}{ |l|| }{$P_{busy}$} & CBD$^3$ & $\overline{P_{d}}$ & CBD$^3$ & $\overline{P_{d}}$ & - & 1 & - & \multicolumn{1}{ |l|| }{1} & $P_{busy} \overline{P_{d}}^2$ & $2E_{s}$ \\ \cline{1-12}

\multicolumn{1}{ |c|| }{S8} & \multicolumn{1}{ |l|| }{$P_{busy}$} & CBD$^3$ & $\overline{P_{d}}$ & MD$^4$ & $\overline{P_{md}}$ & - & 1 & - & \multicolumn{1}{ |l|| }{1} & $P_{busy} \overline{P_{d}}$ $\overline{P_{md}}$ & $2E_{s}$ \\ \cline{1-12}

\multicolumn{1}{ |c|| }{S9} & \multicolumn{1}{ |l|| }{$P_{busy}$} & MD$^4$ & $\overline{P_{md}}$ & CBD$^3$ & $\overline{P_{d}}$ & - & 1 & - & \multicolumn{1}{ |l|| }{1} & $P_{busy} \overline{P_{d}}$ $\overline{P_{md}}$ & $2E_{s}$ \\ \cline{1-12}

\multicolumn{1}{ |c|| }{S10} & \multicolumn{1}{ |l|| }{$P_{busy}$} & MD$^4$ & $\overline{P_{md}}$ & MD$^4$ & $\overline{P_{md}}$ & - & 1 & - & \multicolumn{1}{ |l|| }{1} & $P_{busy} \overline{P_{md}}^2$ & $2E_{s}+E_t+E_r$ \\ \cline{1-12}

\hline 

\hline
\multicolumn{12}{ l }{$^1$False Alarm $||$ $^2$Correct Vacant Detection $||$ $^3$Correct Busy Detection $||$ $^4$Mis-Detection $||$ $^5$At least 1 PU arrives within duration $\tau_t$}  \\
\hline

\end{tabular}
}
\egroup
  \label{ScenarioSOP}
\end{table*}

\subsection{Spectrum Utilization Scenarios}
\label{UtilizationofSpectrumOpportunities}
The spectrum holes are utilized opportunistically by SUs. Without loss of generality, MAC level scheduling between SUs is assumed to be collision-free, which can be enabled for a communicating SU pair via Guaranteed Time Slots through MAC layer protocols such as IEEE 802.15.4 \cite{mac.IEEE}. Additionally, as mentioned in Section \ref{SystemModel}, a central entity or cluster heads can ensure SUs receive customized licensed channel lists for access.

For a successful transmission to take place among two one-hop neighbors, four conditions need to be met in total:

\begin{itemize}
\item The first two conditions are on finding a Spectrum Opportunity (SO) through reliable spectrum sensing on both the transmitter and the receiver sides, and are referred to as conditions for Successful Spectrum Access (SSA).
\item The second two conditions cover the cases that throughout the transmission of the frame in $\tau_t$, no PUs arrive on the channel and no bit errors occur on the transmitted frame, and referred to as conditions for Successful Frame Transmission (SFT).
\end{itemize}

Different events in these four conditions bring about 10 scenarios in total, which are summarized in Table \ref{ScenarioSOP}. Within notations from Table \ref{ScenarioSOP}, it should be noted that $\overline{P_{fa}}=P_{fa}/P_{idle}$, $\overline{P_{v}}=P_{v}/P_{idle}=(P_{idle}-P_{fa})/P_{idle}$, $\overline{P_{d}}=P_{d}/P_{busy}$ and $\overline{P_{md}}=P_{md}/P_{busy}=(P_{busy}-P_{d})/P_{busy}$.

In the scenarios S1-S6 from Table \ref{ScenarioSOP}, the licensed channel (that the SU transmitter $SU_{TX}$ and receiver $SU_{RX}$ are both sensing) is idle during spectrum sensing, presenting an SO. In this context, the $SU_{TX}$ and $SU_{RX}$ could correctly sense this channel as idle and access it subsequently. However, given imperfections in energy detection based spectrum sensing, there might be cases where they decide not to access this channel despite it being idle, summarized as scenarios S1-S3. These 3 scenarios cover the cases that at least one of the $SU_{TX}$ and $SU_{RX}$ decide the channel is busy, despite the SO, referred to as a False Alarm (FA) case. Consequently, SSA does not take place in scenarios S1-S3, and total energy consumed by the $SU_{TX}$ and $SU_{RX}$ is $2E_s$, for spectrum sensing operations. 

\begin{figure}[!ht]
\centering
\includegraphics[width=0.65\columnwidth]{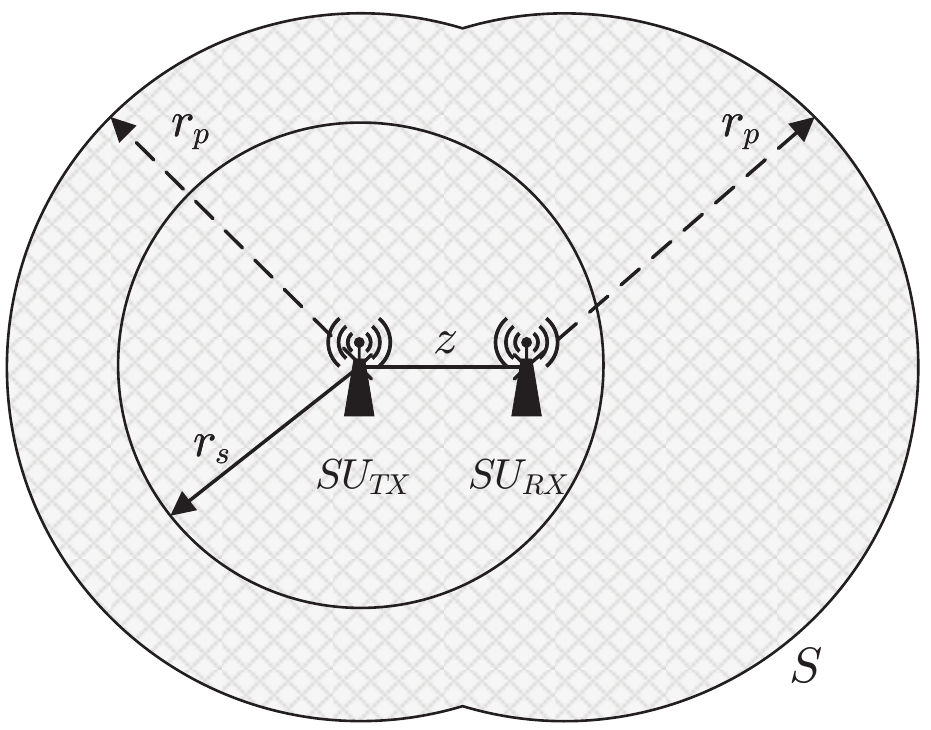}
\caption{Illustration of guardring concept.}
\label{guardring}
\end{figure}

In scenarios S4-S6, the $SU_{TX}$ and $SU_{RX}$ correctly identify the SO and achieve SSA. What differs in between these 3 cases are the fulfillment of two conditions for SFT. First condition is regarding the potential arrival of PU(s) during a time interval of $\tau_t$ on the channel that the $SU_{TX}$ and $SU_{RX}$ previously accessed. Any potential PU in the shaded area around the $SU_{TX}$ and $SU_{RX}$ illustrated in Fig. \ref{guardring} should remain silent during $\tau_t$ so that no PU-SU collision occurs. With the underlying assumption that PU transmission range $r_p$ is greater than SU transmission range $r_s$, no PU arrival probability during $\tau_t$ is expressed as
\begin{equation}
\begin{split}
P_{np} & =\sum_{k=0}^{\infty} \frac{e^{-\rho_{p}S}(\rho_{p} S)^{k}}{k!} (e^{-\tau_{t} \beta})^k \\ 
&= e^{-\rho_{p}S} \Bigg(\sum_{k=0}^{\infty} \frac{(\rho_{p} S e^{-\tau_{t} \beta} )^{k}}{k!}\Bigg)= e^{-\rho_{p}S (1-e^{-\tau_{t} \beta})}, \label{P_np}
\end{split}
\end{equation}
where the first term in (\ref{P_np}) is the probability of having $k$ PUs in the area $S$ illustrated in Fig. \ref{guardring} according to Poisson distribution, and the second term is the probability that all of these $k$ PUs remain silent at least for a duration of $\tau_t$ as per the PU activity model from Section \ref{SystemModel}. The area $S$ between the $SU_{TX}$ and $SU_{RX}$ that are $z$ apart can be defined as 
\begin{equation}
S(z,r_p)=2\pi r_p^2 - 2r_p cos^{-1}\big(\frac{z}{2r_p}\big) + \frac{z}{2}\sqrt{4r_p^2-z^2}.
\label{eq:S}
\end{equation}

The second condition to achieve SFT once SSA is achieved is that all bits of the transmitted frame are received correctly on $SU_{RX}$ side. Under unreliable radio link model \cite{unreliablelink.gorce}, the probability that all the bits in the frame are received successfully can be written as 
\begin{equation}
P_{rel}= (1-BER(\gamma))^\zeta, \label{P_rel}
\end{equation}
where $BER$ is the bit error rate (BER) to evaluate the unreliability of the link, $\gamma$ is the signal-to-noise ratio in the receiver, $\zeta = \tau_t R = \tau_t B \log_2(1+\gamma)$ is the packet size sent during $\tau_t$ in bits with theoretically achievable rate $R$. The BER can be written as $BER(\gamma)= 0.5\ erfc(\sqrt{k \gamma})$ where $erfc(x)=\frac{2}{\sqrt{\pi}} \int_{\sqrt{x}}^{\infty} exp(-u^2)du$ is the complementary error function \cite{ber.zhang}. The channel is assumed to be an AWGN channel, and $k$ depends on the modulation type used. If the modulation type is BPSK, then $k=1$.

Under these conditions, as can be followed from Table \ref{ScenarioSOP}, only scenario S6 refers to SSA followed by SFT. The probability of this scenario can be calculated as follows:
\begin{equation}
\begin{split}
P_{S6} & = P_{v}^2P_{np}P_{rel} \\ 
&= (P_{idle}-P_{fa})^2 (e^{-\rho_{p}S (1-e^{-\tau_{t} \beta})})(1-BER(\gamma))^\zeta
\end{split}
\end{equation}
In scenarios S7-S10, the sensed licensed channel is busy when the $SU_{TX}$ and $SU_{RX}$ start their spectrum sensing, hence there is no SO available. In scenarios S7-S9, at least one of the $SU_{TX}$ and $SU_{RX}$ correctly detects this channel as busy. Consequently, both nodes carry out spectrum handoff to the next channel in their list and total energy consumed by them in that slot is $2E_s$, for spectrum sensing operations. However, in scenario S10, the $SU_{TX}$ and $SU_{RX}$ both incorrectly sense the channel as idle, which in turn triggers the SU frame transmission that will not result in an SFT. The probability of this scenario is
\begin{equation}
P_{S10} = P_{md}^2 = \Bigg( P_{busy} \mathcal{Q}\Big(\frac{2\tau_s B \big(\sigma_p^2+\sigma_n^2)-\delta}{\sqrt{4 \tau_s B (\sigma_p^2+\sigma_n^2)^2}}\Big)\Bigg)^2,
\end{equation}
where $P_{md}$ is the mis-detection probability by an SU. Total energy consumed in that slot for S10 is $2E_s+E_t+E_r$, as $SU_{TX}$ and $SU_{RX}$ try to communicate, despite no SO being available, due to double mis-detection on both sides.

\begin{table}[t]
\centering
\scriptsize
    \caption{{Probability mass function of random variable $T$.}}
	\def\arraystretch{1.6}
  \begin{tabular}{ |c | c |}

    \hline
    \multicolumn{1}{ |c |}{\textbf{Number of trials ($t$)}} & \multicolumn{1}{ c| }{\textbf{$P(T=t)$}} \\ \hline
    \hline
    1 & $P_C$ \\ \hline
    2 & $P_C(P_A+P_B)$ \\ \hline
    3 & $P_C(P_A^2+2P_AP_B+P_B^2)$ \\ \hline
    4 & $P_C(P_A^3+3P_A^2P_B+3P_AP_B^2+P_B^3)$ \\ \hline
    $\cdots$ & $\cdots$ \\ \hline

  \end{tabular}
    \label{TrialsPMF}
\end{table}

From a joint SFT and energy consumption perspective, there are three sets of scenarios:
\begin{itemize}
\item Set $A: \{S1, S2, S3, S7, S8, S9\}$, consisting of scenarios with no SFT and an energy consumption of $2E_s$ only.
\item Set $B: \{S4, S5, S10\}$, consisting of scenarios with no SFT and an energy consumption of $2E_s+E_t+E_r$ due to failed transmission.
\item Set $C: \{S6\}$, consisting of the only scenario with SFT and an energy consumption of $2E_s+E_t+E_r$.
\end{itemize}
As part of finding an energy-efficient transmission range $r_s$ and transmission duration $\tau_t$ for SUs, it is of interest to compute the expected energy consumption among an $SU_{TX}$ and $SU_{RX}$ that are $z$ apart. We are specifically interested in the probability mass function (PMF) for the random variable $T$ that denotes number of time slot trials until event $C$ (i.e. scenario $S6$) resulting in an SFT takes place. Let us denote the probabilities of three independent events $A$, $B$, $C$ as $P_A$, $P_B$, $P_C$ respectively. Then the probability mass function for the random variable $T$, which represents the number of trials until SFT is achieved, can be illustrated as in Table \ref{TrialsPMF} and expressed as:
\begin{equation}
P(T=t)=\sum_{k=0}^{t-1} P_C \bigg( {{t-1}\choose{k}} P_A^k P_B^{t-1-k} \bigg).
\end{equation}
The expected energy consumption until SFT between $SU_{TX}$ and $SU_{RX}$ that are $z$ apart is
\begin{equation}
\label{ESFT}
\begin{split}
E\{E_{sft}\}=\sum_{t=1}^{\infty}&\sum_{k=0}^{t-1} P_C \bigg( {{t-1}\choose{k}} P_A^k P_B^{t-1-k} \bigg)  \\ & \times(E_C+kE_A+(t-1-k)E_B),
\end{split}
\end{equation}
where $E_A=2E_s$ and $E_B=E_C=2E_s+E_t+E_r$. Similarly, expected time until SFT is
\begin{equation}
\label{TSFT}
\begin{split}
E\{\tau_{sft}\}=\sum_{t=1}^{\infty}&\sum_{k=0}^{t-1} P_C \bigg( {{t-1}\choose{k}} P_A^k P_B^{t-1-k} \bigg)  \\ & \times((t-k)(\tau_s + \tau_t) + k \tau_s).
\end{split}
\end{equation}

\section{Energy-Efficient $r_s$ and $\tau_t$ for CRSN}
\label{ProblemFormulation}

Our aim is to lay the foundations to investigate energy-efficient homogeneous transmission range $r_s$ and transmissions duration $\tau_t$ numerically for CRSNs in Section  \ref{NumericalResults}. To that end, we first revisit the strong foundations on expected hop progress, denoted as $E\{W\}$, from the source toward sink at each hop \cite{fixed.range.deng}. The assumed routing approach is a greedy forwarding scheme called Least Remaining Distance (LRD), which would guarantee to minimize the remaining distance to the sink at each hop. Secondly, based on expected hop progress toward the sink, we derive the expected hop distance $E\{Z\}$ under the LRD scheme. Using $E\{Z\}$, the expected energy consumption, i.e., $E\{E_{sft}\}$ and time until SFT, i.e., $E\{\tau_{sft}\}$, can be computed using scenario probabilities as in (\ref{ESFT}-\ref{TSFT}). We later use $E\{W\}$, $E\{Z\}$ and $E\{E_{sft}\}$ to numerically compute optimal homogeneous transmission range $r_s^*$ and transmission duration $\tau_t^*$ in Section \ref{NumericalResults}.

\subsection{Expected Hop Progress}
\label{ExpectedHopProgress}

\begin{figure}[!h]
	\centering
	\includegraphics[width=0.85\columnwidth]{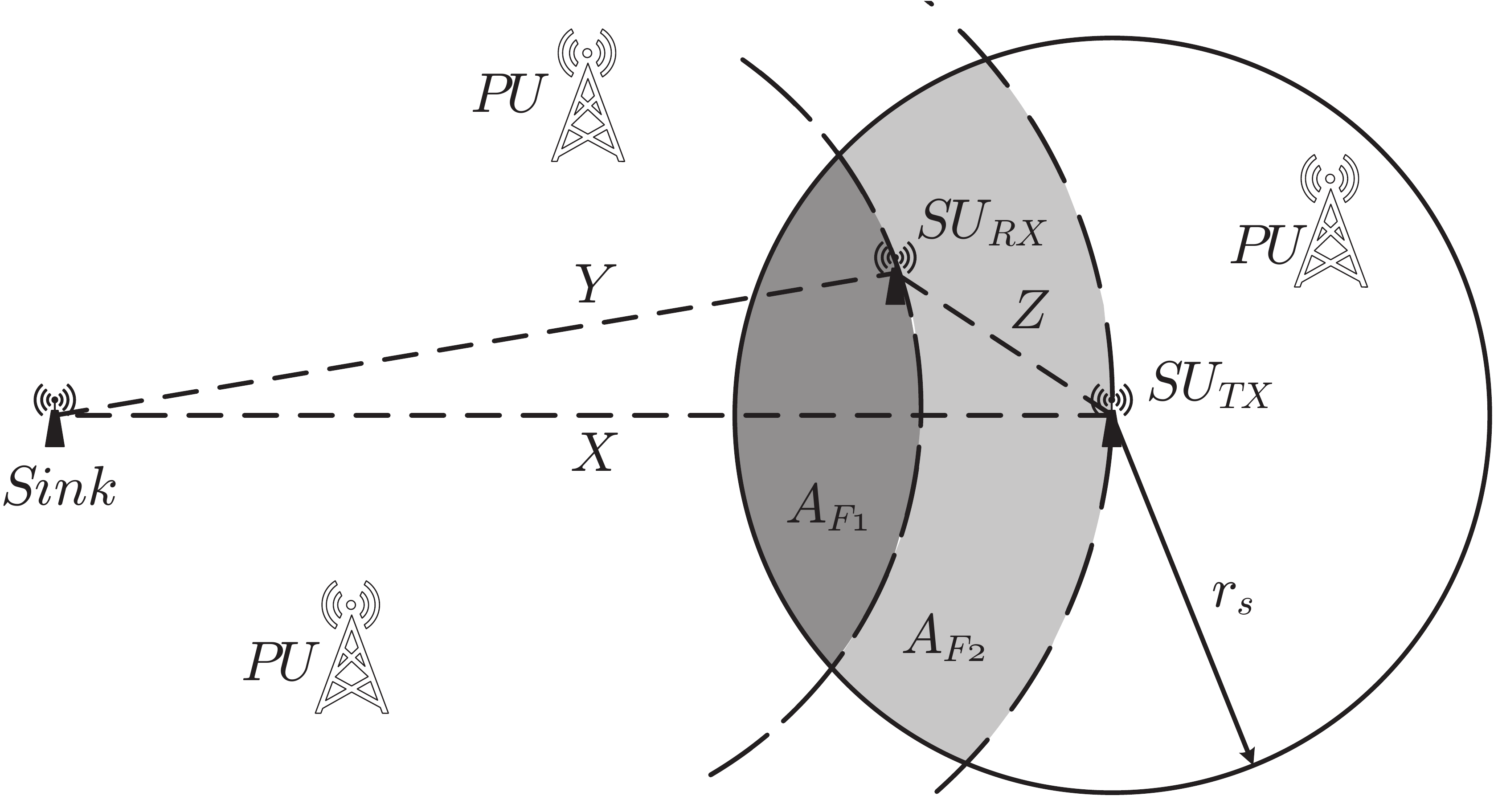}
	\caption{Illustration of forwarding approach.}
	\label{hop_progress}
\end{figure}

To study the expected hop progress toward the sink at each hop, we consider the setup illustrated in Fig. \ref{hop_progress}. Let $SU_{TX}$ be the CRSN node with relevant information that needs to be conveyed to the sink. $SU_{TX}$ has a transmission radius of $r_s$ and would need to trigger multi-hopping toward the sink, unless $r_s<X$, in which case a single hop would suffice. $SU_{TX}$ employs the LRD as the greedy forwarding scheme, choosing its neighbor in the forwarding region that is closest to the sink as the next hop and attempts SSA and SFT subsequently. The forwarding region for $SU_{TX}$ is defined by $A_F=A_{F1} \cup A_{F2}$.

Given DSA and successful communication probabilities derived in detail in Section \ref{UtilizationofSpectrumOpportunities}, $SU_{TX}$ continues its attempts until SFT is achieved. At each hop, the same process is repeated, hence, each hop can be regarded as the first step of the remaining forwarding process toward the sink \cite{fixed.range.de, fixed.range.deng}.

For the setup in Fig. \ref{hop_progress}, assume that $SU_{TX}$ chooses node $SU_{RX}$ as its next hop neighbor according to LRD greedy forwarding \cite{greedy.forwarding.karp}. Per system model characteristics, both SU and PU locations are assumed to be random. Therefore, all the distances between $SU_{TX}$, $SU_{RX}$ and the sink can be characterized by random variables. Let the distances between $SU_{TX}$ and the sink, $SU_{RX}$ and the sink and $SU_{TX}$ and $SU_{RX}$ be represented by the random variables $X$, $Y$ and $Z$, respectively. In this context, $W = X-Y$ is the random variable that represents hop progress. LRD forwarding chooses the neighbor of $SU_{TX}$ in the region $A_F=A_{F1} \cup A_{F2}$ with minimum $Y$, which in turn maximizes $W$. In the example in Fig. \ref{hop_progress}, this neighbor is $SU_{RX}$, with no other neighbor closer to the sink being present in the region $A_{F1}$.

$W$ also depends on the distance between $SU_{TX}$ and the sink. When $X \leq r_s$, $SU_{TX}$ and the sink are one-hop neighbors, the hop progress is then automatically equal to $X$. If $X > r_s$, there must be a neighbor node in the forwarding region for hop progress toward the sink. This condition can be represented by the following random variable:  
\begin{equation}
   F = \left\{
     \begin{array}{ll}
       1 &, \textrm{if there is a neighbor in the forwarding region,} \\
       0 &, \textrm{otherwise}.
     \end{array}
   \right.
\end{equation}
The expected hop progress is described and calculated as in (\ref{eq:expected_hop_progress}), which is studied in detail in \cite{fixed.range.deng}. Since CRSNs have usually high and even ultra-high densities, the following approximations, 
\begin{equation}
\label{eq:approx}
1 - e^{-\rho_s A_{F1}} \approx 1 \ \textrm{and} \ 1 - e^{-\rho_s A_{F}} \approx 1, 
\end{equation}
can be made. Then, $E\{W\}$ expression simplifies to  
\begin{equation}
E\{W\} = r_s - \frac{r_s^3}{3 \Gamma^2}.
\end{equation}

\begin{figure*}[!b]
	\hrulefill
	
	\begin{IEEEeqnarray}{rCl}
		\begin{split}
			\displaystyle
			E\{W|((X \leq r_s) \cup ( X > r_s \cap F=1 ))\} = \int_0^{\infty} Pr\{W>w|((X \leq r_s) \cup ( X > r_s \cap F=1 ))\} dw\\
			= \displaystyle \frac{3 \Gamma^2 r_s - 6 \int_0^{r_s} \int_{r_s}^{\Gamma}x e^{-\rho_s A_{F1}} dx\ dw}{3 (x^2 - 2 \int_{r_s}^\Gamma xe^{-\rho_s A_F}dx)} \\  
		\end{split}
		\label{eq:expected_hop_progress}
	\end{IEEEeqnarray}
\end{figure*} 

\subsection{Expected Hop Distance}
\label{ExpectedHopDistance}
\begin{figure}[!ht]
	\centering
	\includegraphics[width=0.85\columnwidth]{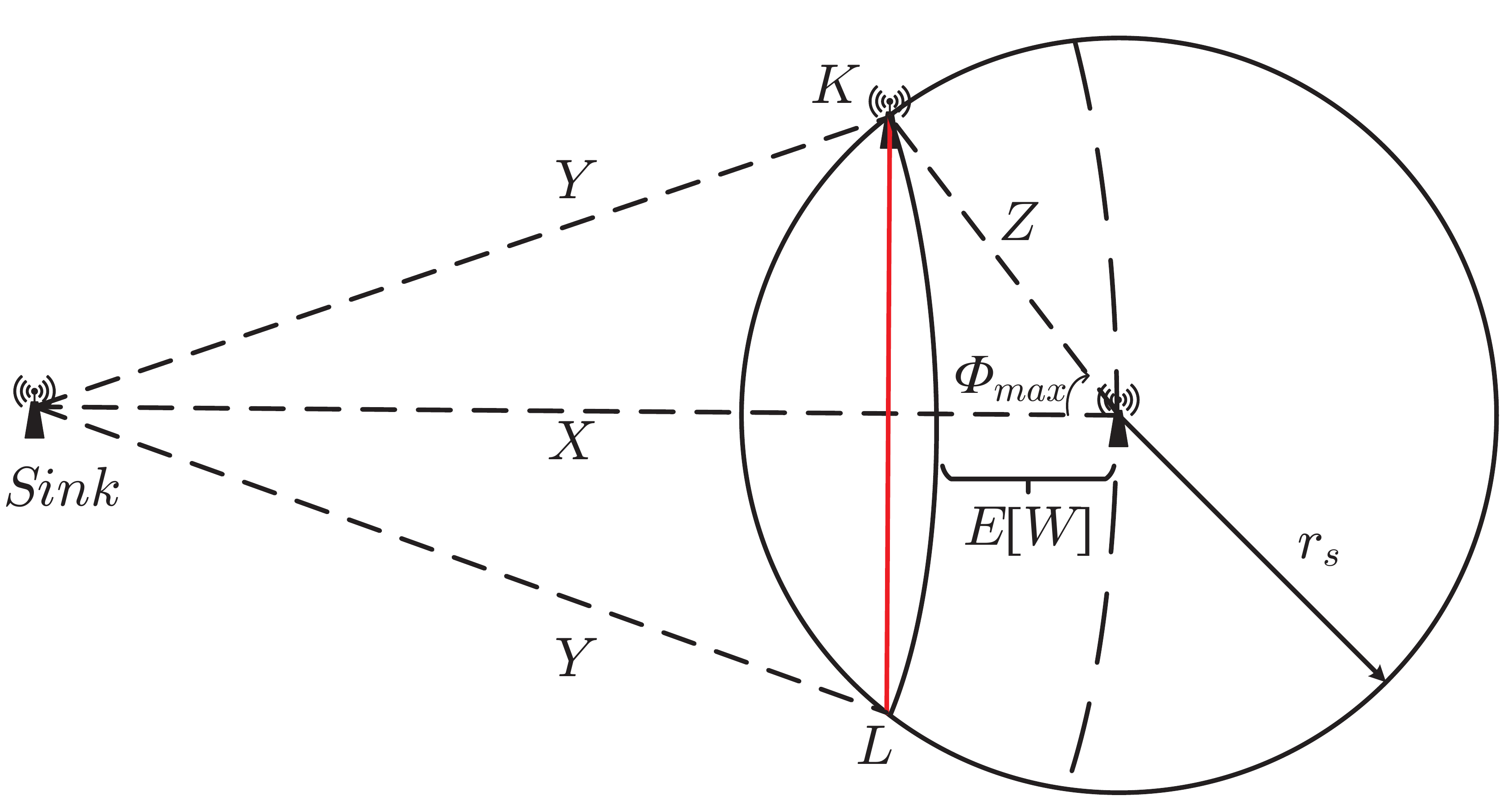}
	\caption{Relation between hop progress and hop distance.}
	\label{hop_distance}
\end{figure}

Although $r_p>r_s$ is strictly assumed, it is also considered that transmission ranges of PUs and SUs are comparable, so that channel availability for SUs has both spatial and temporal variations, as in our previous work \cite{spectrum.opportunity.ozger}. Hence, SO depends on the distance between $SU_{TX}$ and $SU_{RX}$. This distance, depicted as $Z$ in Fig. \ref{hop_distance}, determines the area of the guardring (as per (\ref{eq:S})) where there should be no active PU for a duration of $\tau_t$ after SSA. To that end, $Z$ has a direct impact on the expected energy $E\{E_{sft}\}$ and time $E\{\tau_{sft}\}$ required to achieve SFT.

For a given $E\{W\}=w$, the next hop node $SU_{RX}$ can be uniformly anywhere on the black solid arc that is from point K to point L as seen in Fig. \ref{hop_distance}. We assume linear approximation for the KL arc to decrease the computational burden, which is shown by the solid red line in Fig. \ref{hop_distance}. With this approximation, the angle $\Phi$ between the edges that are $X$ and $Z$ long is uniformly distributed in $[-\Phi_{max}, \Phi_{max}]$. With the aforementioned approximation, for a given $E\{W\}=w$, it is clear that $\Phi_{max}=cos^{-1}(\tfrac{E\{W\}}{r_s})=cos^{-1}(\tfrac{w}{r_s})$. Additionally, the hop distance can be written as $Z=\tfrac{w}{cos(\Phi)}$ for a random point on the red line, with its maximum being equal to $r_s$. The expected hop distance for a given $E\{W\}=w$ is then
\begin{equation} \label{eq:expected_hop_distance}
\begin{split}
 E\{Z|&w\} = \int_{-\Phi_{max}}^{\Phi_{max}} \frac{w}{2 \Phi_{max} cos(\Phi)} d \Phi, \\
 			 &= \frac{w}{2\Phi_{max}} \bigg( ln(tan(\Phi_{max}) + sec(\Phi_{max})) \\
 			 & \;\;\;\;\;\;\; -ln(tan(-\Phi_{max}) + sec(-\Phi_{max}))\bigg).
\end{split}
\end{equation}  
The expected value given in (\ref{eq:expected_hop_distance}) is evaluated and plugged into (\ref{eq:S}) to calculate guardring sizes for different network scenarios in Section \ref{NumericalResults}.

\subsection{Energy-Efficient Transmission Range and Duration}
\label{Energy-EfficientTransmissionRangeDuration}
Based on the derivations for SSA and SFT probabilities from Section \ref{UtilizationofSpectrumOpportunities} and study of expected hop progress toward the sink and hop distance in a greedy forwarding scheme from Sections \ref{ExpectedHopProgress} and \ref{ExpectedHopDistance}, we are ready to formulate the optimization problem. The optimization aims to find ($r_s*$, $\tau_t*$) pair that would maximize the goodput-meter per Joule achieved in the provided CRSN system model. Hence, the optimization problem can be formulated as a reward function $\Lambda$ as
\begin{equation}
\begin{aligned}
& \underset{\displaystyle r_{s}, \tau_{s}}{\text{max}}
& & \Lambda=\dfrac{\dfrac{\tau_t R}{E\{\tau_{sft}\}} \cdot E\{W\}}{E\{E_{sft}\}} \\ 
& \text{subject to}
& & 0 < r_s < r_p, \\
&&& 0 < \tau_t < \tau_{max},\
\end{aligned}
\label{eq:Reward function}
\end{equation}
where the unit for $\Lambda$ is (bits/s)$\times$m per Joule. The maximum transmission duration constraint, $\tau_{max}$, is
\begin{equation}
\tau_{max}=-\frac{1}{\mu}\log\Big(1-\frac{P_{col}}{P_{idle}}\Big)
\label{eq:tau_max}
\end{equation}
where $\mu=max\{\alpha, \beta\}$ and $P_{col}$ is the maximum SU-PU collision probability that a PU can tolerate \cite{optimal.spectrum.sensing.lee}. Thanks to our approximation in \eqref{eq:approx}, \eqref{eq:Reward function} becomes an optimization problem with a quadratic objective function with linear constraints.

\section{Numerical Results}
\label{NumericalResults}
In this section, first of all, we provide the preliminaries on required spectrum sensing duration $\tau_s$ given the desired transmission duration $\tau_t$ and the maximum SU-PU collision probability that a PU can tolerate $P_{col}$ \cite{optimal.spectrum.sensing.lee}. Consequently, behavior of the SFT probability $P_{S6}$ as the underlying driver of the reward function $\Lambda$ given in (\ref{eq:Reward function}), is studied for different ($r_s$, $\tau_t$) scenarios. Moving forward, the reward function $\Lambda$ is evaluated and examined in detail. As part of that, sensitivities of optimal ($r_{s}^*$, $\tau_t^*$, $\Lambda^*$) values with respect to changes in 9 critical network parameters are tabulated in detail. As a last step, how optimal ($r_{s}^*$, $\tau_t^*$, $\Lambda^*$) values respond to combined changes in path loss exponent $\kappa$, PU death rate $\alpha$ and birth rate $\beta$ are examined. Employed parameters and their values throughout all following analyses, unless otherwise stated, are provided in Table \ref{table:Parameters}.

\begin{table}[!t]
\centering
\scriptsize
\def\arraystretch{1.6}
\caption{Parameters and their assumed values  {from \cite{fixed.range.deng, fixed.range.chen, optimal.spectrum.sensing.lee}}.}
\begin{tabular}{ l  l  r } 
\hline
\textbf{Parameter} & \textbf{Symbol} & \textbf{Value}   \\
\hline\hline
PU birth rate & $\beta$ & $3$ / s  \\  \hline
PU death rate & $\alpha$ & $3$ / s  \\  \hline
PU guardring radius & $r_p$ & $200$ m  \\  \hline
Path loss exponent & $\kappa$ & $2.5$ \\ \hline
Reference SU SNR & $\gamma_0$ & $20$ dB \\ \hline
PU signal vs. noise variance & $\gamma_p = \sigma_{p}^2/\sigma_{n}^2$ & 10 \\ \hline
Receiver noise figure & $F_{rx}$ & $12.589$\\ \hline
Thermal noise & $N_0$ & $4.17 \times 10^{-21}$ W/Hz\\ \hline
Bandwidth & $B$ & $10$ KHz  \\ \hline
Signal wavelength & $\lambda$ & $0.125$ m \\  \hline 
Amplifier efficiency & $\eta_{amp}$ & $0.2$  \\  \hline 
Antenna gain & $G_{a}$ & $0.01$  \\  \hline 
TX circuit power consumption & $P_{elec}$ & $3.63$ mW \\  \hline
Spectrum sensing power cons. & $P_{s}$ & $700$ mW \\  \hline 
RX total power consumption & $P_{rx}$ & $11.13$ mW \\  \hline 
Network diameter & $\Gamma$ & $1000$ m \\  \hline
\end{tabular}
\label{table:Parameters}
\end{table}

\subsection{Preliminaries on $\tau_t$ and $\tau_s$}
\label{subsection:tau_t and tau_s}
Transmission duration $\tau_t$ is one of the two decision variables ($r_s$, $\tau_t$) for the reward function $\Lambda$ in (\ref{eq:Reward function}) that is intended to be maximized. Sensing duration $\tau_s$ also directly affects SFT probability $P_{S6}$ in a slot for the given system model, yet cannot be chosen arbitrarily. In fact, $\tau_s$ is dependent on the desired $\tau_t$ value, driven by the maximum SU-PU collision probability that a PU can tolerate $P_{col}$, hence is not a decision variable in our problem formulation.

\begin{figure*}[!ht]
	
	\begin{subfigure}[h]{0.5\textwidth}
		\centering
		\includegraphics[width=0.7\textwidth]{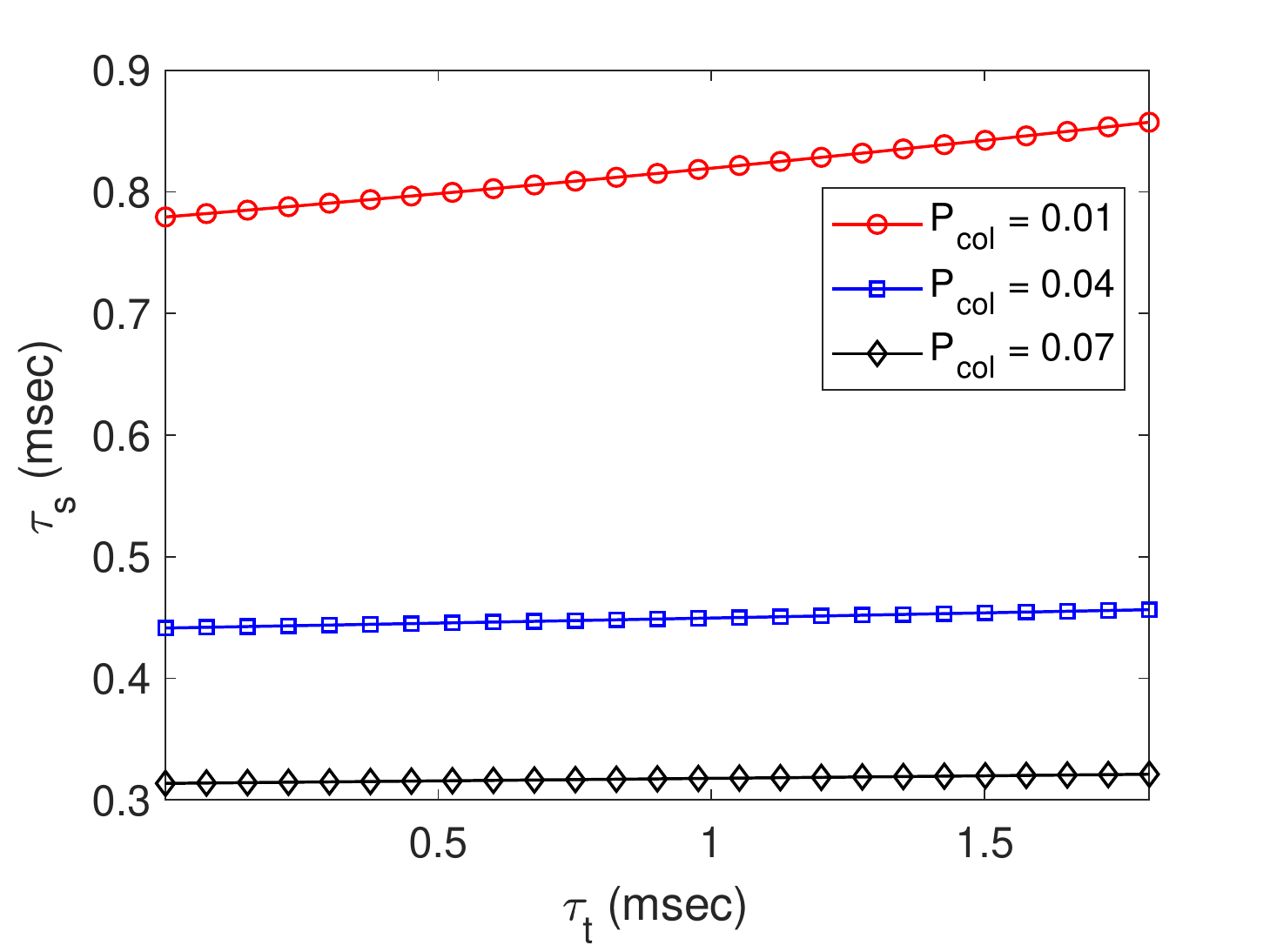}
		\caption{}\label{fig:Figure_5a}
	\end{subfigure}%
	\begin{subfigure}[h]{0.5\textwidth}
		\centering
		\includegraphics[width=0.7\textwidth]{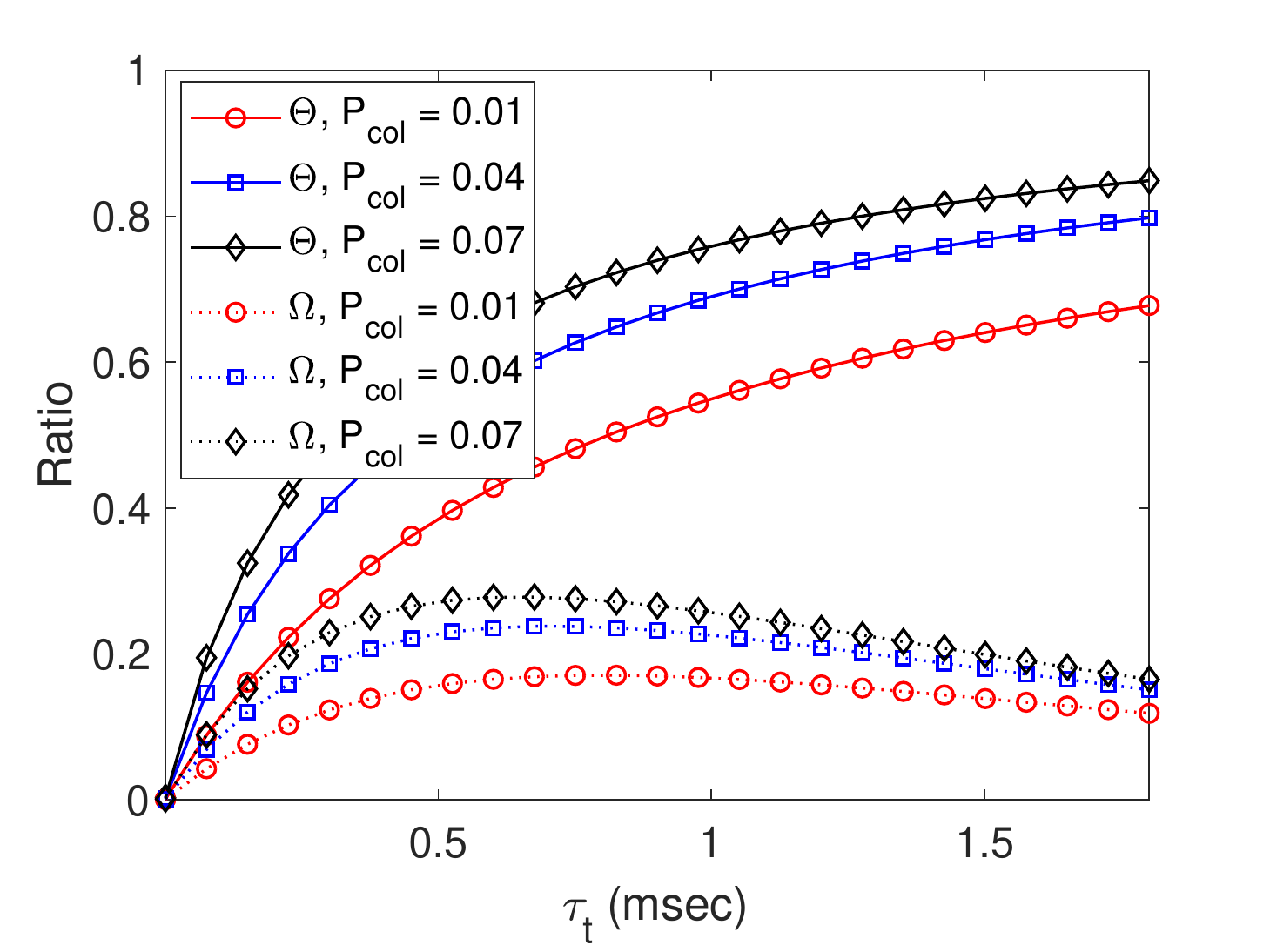}
		\caption{}\label{fig:Figure_5b}
	\end{subfigure}%
	
	\caption{(a) Required spectrum sensing duration $\tau_s$ vs. $\tau_t$ for different $P_{col}$ cases; (b) Comparison of sensing efficiency (defined as $\Theta=\frac{\tau_s}{\tau_s+\tau_t}$) and SFT efficiency (defined as $\Omega=\frac{\tau_t}{E\{\tau_{sft}\}}$ \cite{spectrum.opportunity.bicen} for different $P_{col}$ cases.}  
	\label{fig:Figure_5}
\end{figure*}

For the optimization purposes in this paper, $\tau_s$-$\tau_t$ relation from \cite{optimal.spectrum.sensing.lee} has been adopted. More specifically, for given $\alpha$, $\beta$, $\tau_t$ and $P_{col}$ values, the maximum value of the false alarm probability allowed as per $P_{col}$ requirement is
\begin{equation}
P_{fa} = P_{idle}P_{busy} - P_{idle}P_{busy}\bigg(1-\frac{P_{col}}{P_{idle}}\bigg)e^{\mu \tau_t}
\end{equation}
and $\tau_s$ is picked such that mis-detection and false alarm probabilities are equal, i.e. $P_{fa}=P_{md}$. This serves to balance interference caused upon PUs through mis-detection and lost spectrum opportunities through false alarm. With this approach, $\tau_s$ is given by 
\begin{equation}
\tau_s = \frac{1}{B \gamma_{p}^2} \bigg(Q^{-1}(\frac{P_{fa}}{P_{idle}})+(\gamma_p+1)Q^{-1}(\frac{P_{fa}}{P_{busy}})\bigg)^2.
\end{equation}
Required $\tau_s$ for different $P_{col}$ constraints are depicted in Fig. \ref{fig:Figure_5}(a). Inherently,  required $\tau_s$ that satisfies $P_{fa}=P_{md}$ is prolonged as $P_{col}$ constraint gets stricter. Additionally, with higher $P_{col}$, slope of the $\tau_t$ vs. $\tau_s$ increases.

As per this $\tau_s$-$\tau_t$ relation, it is of interest to reveal how much of the spectrum is intended and used for communications across time. Two metrics can be employed for that purpose. First one is the spectrum efficiency, defined as the intended ratio for communications in a slot, i.e. $\Theta=\frac{\tau_s}{\tau_s+\tau_t}$ \cite{optimal.spectrum.sensing.lee}. Second one is the SFT efficiency, defined as the ratio of time used for communications to actual time spent until SFT is achieved, i.e. $\Omega=\frac{\tau_t}{E\{\tau_{sft}\}}$. Fig. \ref{fig:Figure_5}(b) illustrates both metrics for different $P_{col}$ cases. As seen in Fig. \ref{fig:Figure_5}(b), $\Theta$ monotonically increases in the given $\tau_t$ interval, whereas $\Omega$ peaks around $0.6$-$0.8$ msec interval, depending on the case for $P_{col}$. Although neither $\Theta$ nor $\Omega$ considers energy efficiency, $\Omega$ provides a better indication for likely optimal $\tau_{t}^*$ values for the problem here, given the $(\tau_t R)/E\{\tau_{sft}\}$ term in (\ref{eq:Reward function}). Optimal $\tau_t$ values from $\Omega$ perspective, i.e. $\tau_{t}^{*,\Omega}$, for $P_{col}=0.07$ is $\tau_{t}^{*,\Omega}=0.63$ msec, for $P_{col}=0.04$ is $\tau_{t}^{*,\Omega}=0.72$ msec and for $P_{col}=0.01$ is $\tau_{t}^{*,\Omega}=0.78$ msec. As stricter $P_{col}$ constraints drive $\tau_s$ significantly, it is logical that $\tau_{t}^{*,\Omega}$ follows this upward trend.

Consequently, in the following analyses, SFT efficiency $\Omega$ and suitable values for $\tau_s$-$\tau_t$ thereof will be considered to optimize the reward function in (\ref{eq:Reward function}).

\subsection{Behavior of SFT probability $P_{S6}$}
\label{subsection:P_S6}
\begin{figure*}[!ht]
\centering
	\begin{subfigure}[h]{0.31\textwidth}
		\includegraphics[width=\textwidth]{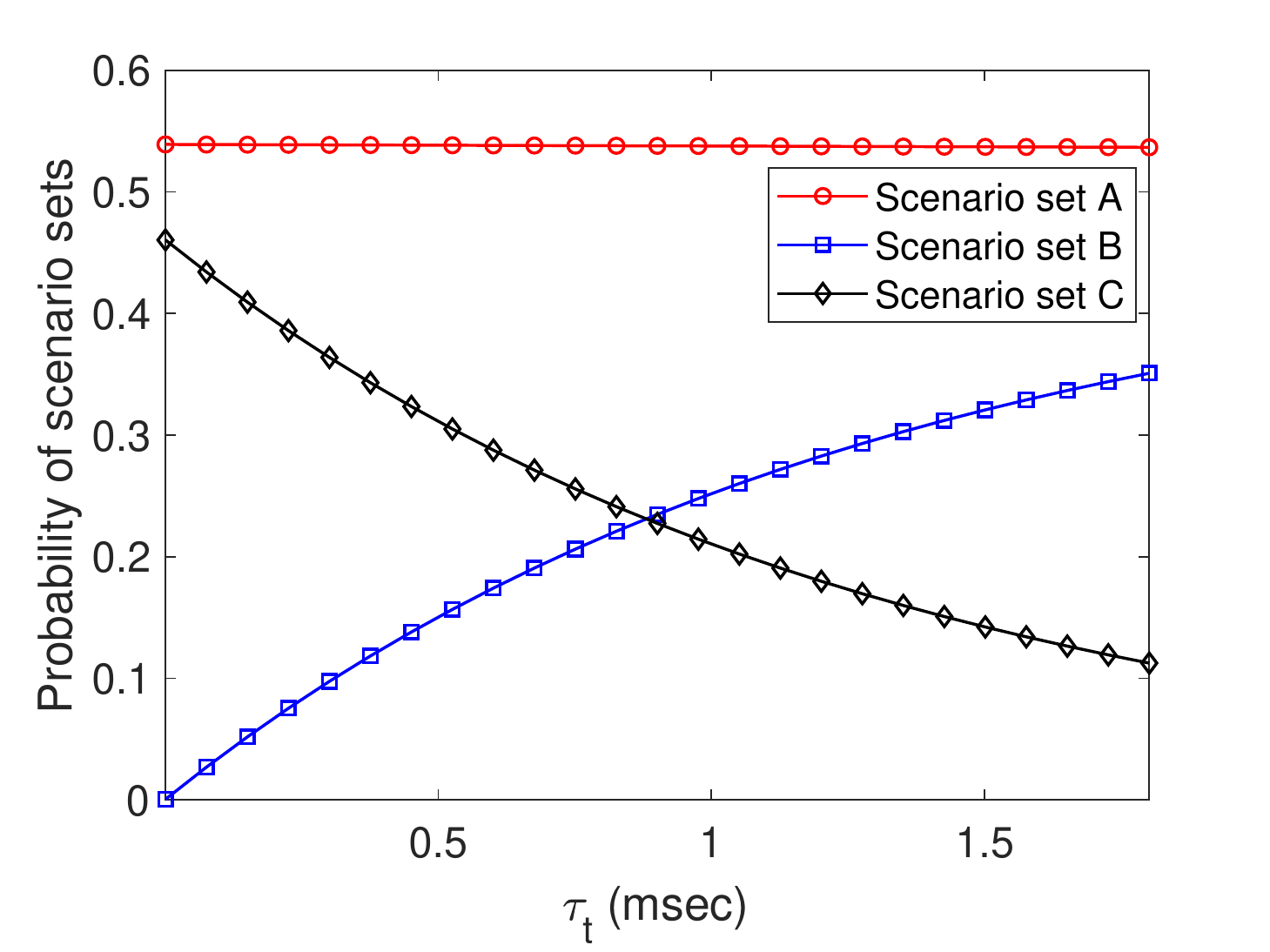}
		\caption{}\label{fig:Figure_6a}
	\end{subfigure}%
	\begin{subfigure}[h]{0.31\textwidth}
		\includegraphics[width=\textwidth]{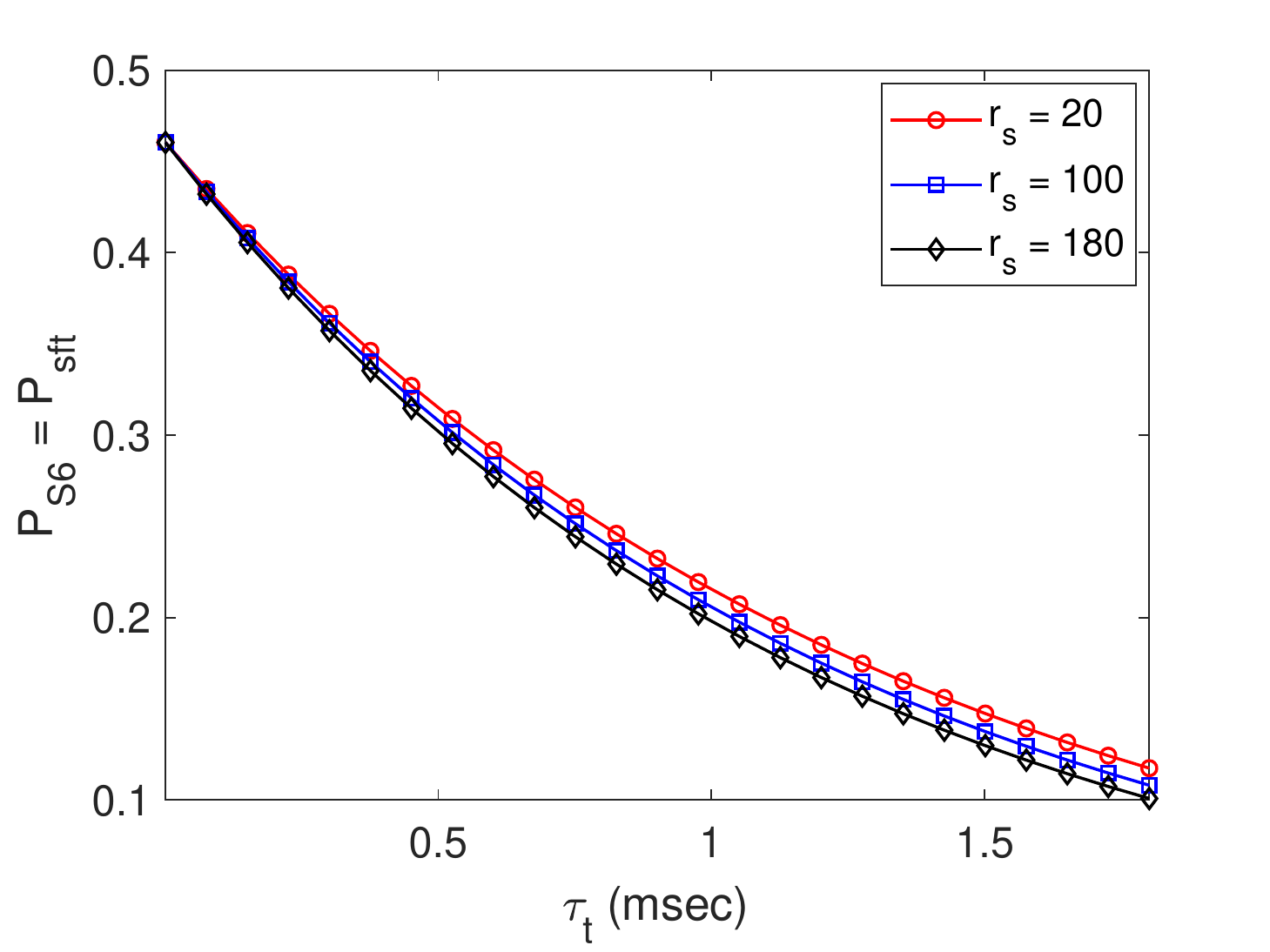}
		\caption{}\label{fig:Figure_6b}
	\end{subfigure}%
	\begin{subfigure}[h]{0.31\textwidth}
		\includegraphics[width=\textwidth]{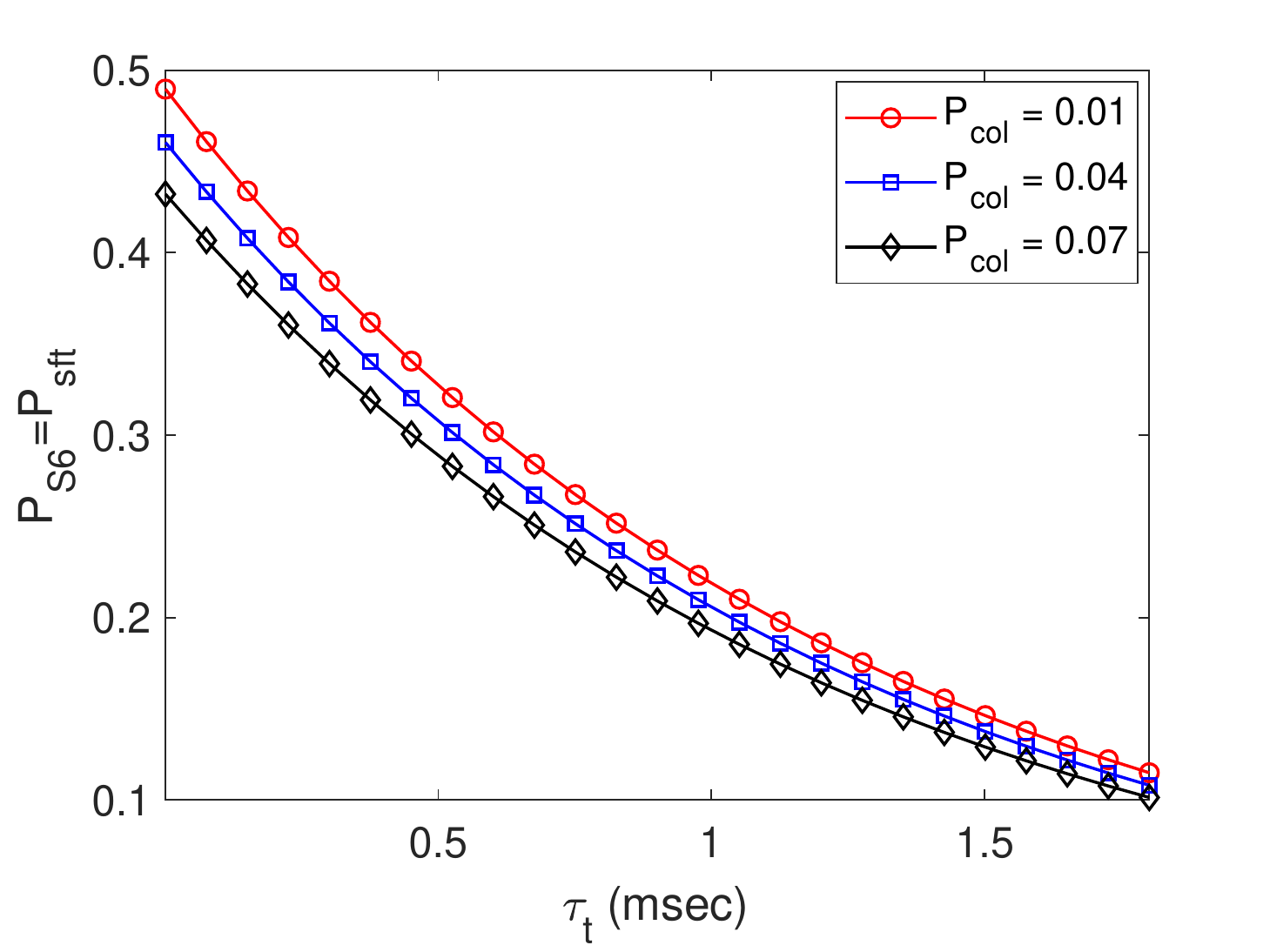}
		\caption{}\label{fig:Figure_6c}
	\end{subfigure}\\%
	
	\caption{(a) Probabilities of scenario sets A, B and C vs. $\tau_t$; (b) Probability of success scenario $P_{S6}$ vs. $\tau_t$ for different $r_s$ cases; (c) Probability of success scenario $P_{S6}$ vs. $\tau_t$ for different $P_{col}$ cases.}  
	\label{fig:Figure_6}

\end{figure*}

As part of the optimization process of the reward function $\Lambda$ in (\ref{eq:Reward function}), it is important to reveal how the probability of SFT, i.e. $P_{S6}$ behaves and varies with critical parameters. Fig. \ref{fig:Figure_6}(a) depicts the probabilities of scenario sets $A$, $B$ and $C$ (which were previously defined in Section \ref{UtilizationofSpectrumOpportunities}) with respect to $\tau_t$, using parameter values defined in Table \ref{table:Parameters}. Accordingly, the probability of set $A$, which includes scenarios related to no SSA and is denoted by $P_A$, is relatively flat with changing $\tau_t$. This result is natural, given the fact that $P_A$ includes two types of events:
\begin{itemize}
\item Sensed channel is idle, yet at least one of the SU TX and RX sense the channel as occupied.
\item Sensed channel is busy, and at least one of the SU TX and RX sense the channel as occupied.
\end{itemize}
The probabilities of both of these event types depend on spectrum sensing duration $\tau_s$, which is also relatively flat with changing $\tau_t$, as previously illustrated in Fig. \ref{fig:Figure_5}(a). On the other hand, the probability of SFT, i.e. $P_{C}=P_{S6}$ decays quickly toward null with increasing $\tau_t$. To that end, $P_{S6}$ and hence reward function $\Lambda$ in (\ref{eq:Reward function}) are significantly sensitive to changes in $\tau_t$.

Fig. \ref{fig:Figure_6}(b) breaks-down the same $P_{S6}$ further by different $r_s$ cases with respect to $\tau_t$. Consequently, the effect of varying $r_s$ on $P_{S6}$ is insignificant when $\tau_t$ is low, yet suddenly becomes material with increasing $\tau_t$. In fact, increasing $r_s$ from $20$ m  to $180$ m reduces $P_{S6}$ only by $1\%$ when $\tau_t=0.5$ msec; yet same $r_s$ change reduces $P_{S6}$ by $14\%$ when $\tau_t=1.8$ msec. To that end, effect of $r_s$ and $\tau_t$ changes on $P_{S6}$ are correlated and requires further deep-dive.

Fig. \ref{fig:Figure_6}(c) details the combined effect of $P_{col}$ and $\tau_t$ on $P_{S6}$. A stricter $P_{col}$ requirement brings a lower $P_{fa}=P_{md}$, hence SFT probability $P_{S6}$ is higher for $P_{col}=0.01$ compared to when $P_{col}=0.07$ as in Fig. \ref{fig:Figure_6}(c). $P_{S6}$ reduces by $\sim 10\%$ as $P_{col}$ is relaxed from $0.01$ to $0.07$, with very little variation as $\tau_t$ is increased. To that end, effect of $P_{col}$ and $\tau_t$ changes on $P_{S6}$ are uncorrelated. Additionally, although a stricter $P_{col}$ brings a higher $P_{S6}$ in each slot, it still results in a lower SFT efficiency $\Omega$ as previously revealed in Fig. \ref{fig:Figure_5}(b).

\subsection{Reward Function $\Lambda$ With Respect to $r_s$ and $\tau_t$}
\label{subsection:Reward function illustration}
\begin{figure}[!h]
	\centering
	\includegraphics[width=0.9\columnwidth]{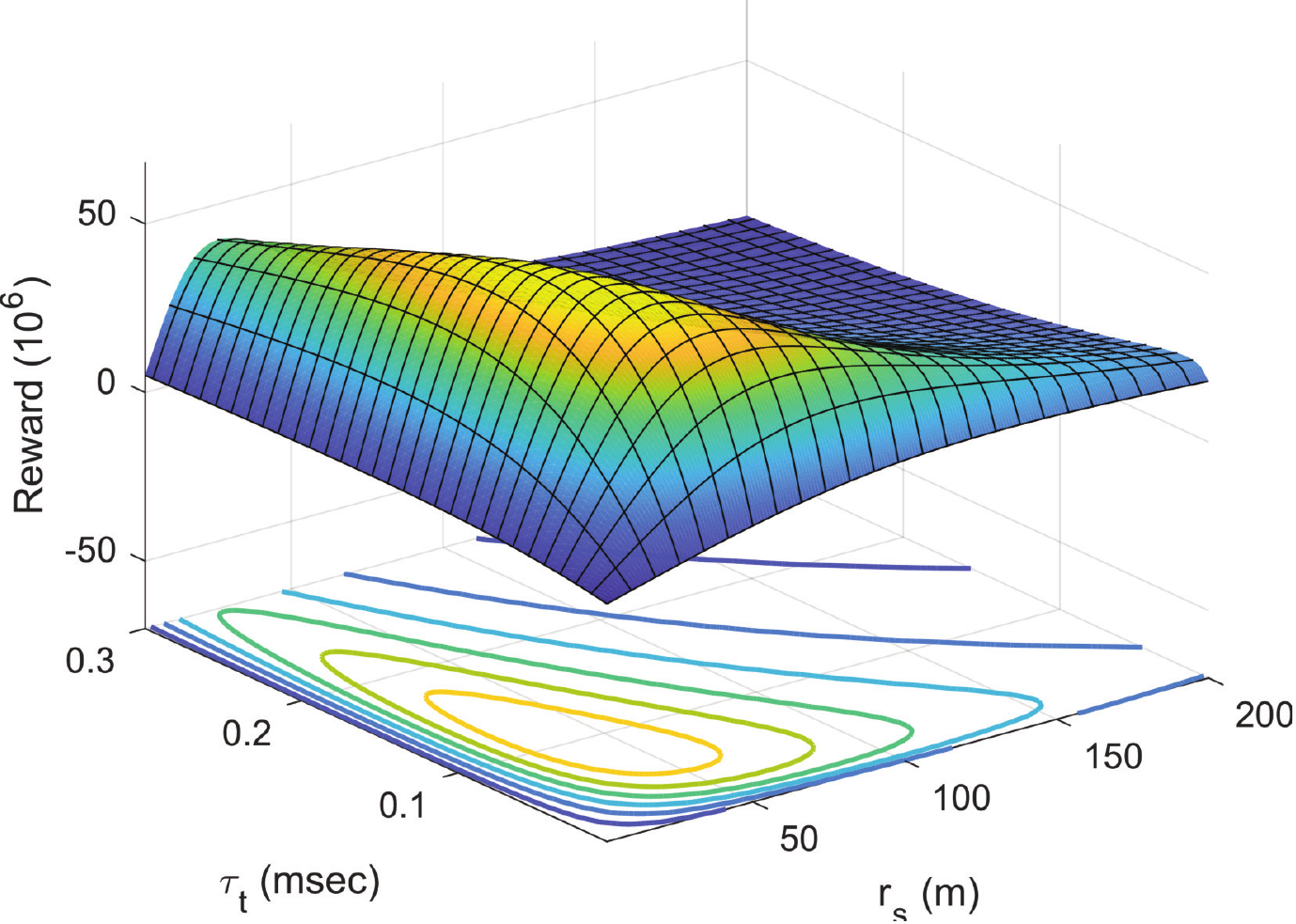}
	\caption{Illustration of the reward function for values in Table \ref{table:Parameters}.}
	\label{fig:Figure_7}
\end{figure}

After the preliminary on $\tau_s$-$\tau_t$ relation, and how $P_{S6}$ reacts to changes in $\tau_s$, $\tau_t$ and $P_{col}$; it is time to construct the reward function $\Lambda$. Fig. \ref{fig:Figure_7} provides a sample illustration of $\Lambda$ for the parameter values given in Table \ref{table:Parameters}. The reward function $\Lambda$ is smooth and well defined for all $r_s<r_p$ and $\tau_t<\tau_{max}$, as seen in Fig. \ref{fig:Figure_7}. For this sample illustration, optimal ($r_{s}^*$, $\tau_{t}^*$, $\Lambda^*$) trio is ($40$ m, $0.1$ msec, $6.8 \times 10^{7}$ (bps$\times$m)/Joule). We examine in detail how ($r_{s}^*$, $\tau_{t}^*$, $\Lambda^*$) trio responds to changes in critical network parameters in Section \ref{sec:Sensitivity subsection}.

\begin{table}[!t]
\scriptsize
\centering
\def\arraystretch{1.2}
\caption{Changes to terms of the reward function $\Lambda$ in (\ref{eq:Reward function}) when only individual parameters are increased.}
\begin{tabular}{c | c c c c}
\hline
\textbf{Event} & $\mathbf{E\{\tau_{sft}\}}$ & $\mathbf{E\{E_{sft}\}}$ & $\mathbf{E\{W\}}$ & $\mathbf{R}$ \\
\hline
$\kappa$ $\uparrow$ & $\leftrightarrow$ & $\uparrow$ & $\leftrightarrow$ & $\downarrow$ \\
\hline
$\beta$ $\uparrow$ & $\uparrow$ & $\uparrow$ & $\leftrightarrow$ & $\leftrightarrow$ \\
\hline
$\alpha$ $\uparrow$ & $\downarrow$ & $\downarrow$ & $\leftrightarrow$ & $\leftrightarrow$ \\
\hline
$\rho_p$ $\uparrow$ & $\uparrow$ & $\uparrow$ & $\leftrightarrow$ & $\leftrightarrow$ \\
\hline
$r_p$ $\uparrow$ & $\uparrow$ & $\uparrow$ & $\leftrightarrow$ & $\leftrightarrow$ \\
\hline
$\gamma_0$ $\uparrow$ & $\leftrightarrow$ & $\uparrow$ & $\leftrightarrow$ & $\uparrow$ \\
\hline
$P_s$ $\uparrow$ & $\leftrightarrow$ & $\uparrow$ & $\leftrightarrow$ & $\leftrightarrow$ \\
\hline
$P_{col}$ $\uparrow$ & $\downarrow$ & $\downarrow$ & $\leftrightarrow$ & $\leftrightarrow$ \\
\hline
$\sigma_p^2/\sigma_n^2$ $\uparrow$ & $\downarrow$ & $\downarrow$ & $\leftrightarrow$ & $\leftrightarrow$ \\
\hline
\multicolumn{5}{ c }{$\uparrow$: Increases $\|$ $\downarrow$: Decreases $\|$   $\leftrightarrow$: No change observed}
\end{tabular}
\label{table:Sensitivity sidetable}
\end{table}

\subsection{Sensitivity of Optimal $r_{s}^*$, $\tau_{t}^*$ and $\Lambda^*$}
\label{sec:Sensitivity subsection}
Assuming parameter base values and ($r_{s}^*$, $\tau_{t}^*$) are kept, Table \ref{table:Sensitivity sidetable} reveals how individual changes in parameters, all else being equal, impact each term on the reward function $\Lambda^*$.

On the other hand, Table \ref{table:Sensitivity} presents the sensitivity of optimal ($r_{s}^*$, $\tau_{t}^*$, $\Lambda^*$) to individual changes in 9 critical network parameters, ceteris paribus. The base values for each parameter, from which the sensitivities are tested, are provided in second column of Table \ref{table:Sensitivity}. Third and fourth columns together indicate how sensitive optimal ($r_{s}^*$, $\tau_{t}^*$, $\Lambda^*$) trio are to changes in the corresponding network parameter. Next paragraphs detail out the impact of each parameter on network design and provides the reasoning thereof.

\begin{table*}[!htbp]
	\scriptsize
	\centering
	\caption{{Sensitivity of optimal $r_s^*$, $\tau_t^*$ and $\Lambda^*$ values with respect to changes in critical network parameters.}}
	\bgroup
	\def\arraystretch{1.4}
	\resizebox{\textwidth}{!}{%
	\begin{tabularx}{\textwidth}{XXX X XXXX  XXXX }
		\cline{5-12}
		& & & & \multicolumn{8}{ |c| }{\textbf{ Change in decision variables and reward when parameter is changed by...}} \\
		\cline{1-12}

		\multicolumn{1}{ |c|  }{\textbf{\scriptsize Parameter}} & \multicolumn{1}{ c|  }{\textbf{Base value}} & \multicolumn{1}{ c  }{\textbf{Variable}} & \multicolumn{1}{ |c  }{\textbf{Sensitivity}} & \multicolumn{1}{ |c|  }{$-50\%$} & \multicolumn{1}{ c| }{$-25\%$} & \multicolumn{1}{ c|  }{$-10\%$} & \multicolumn{1}{ c|  }{$-5\%$} & \multicolumn{1}{ c|  }{$+5\%$} & \multicolumn{1}{ c|  }{$+10\%$} & \multicolumn{1}{ c|  }{$+25\%$} & \multicolumn{1}{ c|  }{$+50\%$} \\ \cline{1-12}

		&  & \multicolumn{1}{ |c }{$r_{s}^*$} & \multicolumn{1}{ c }{\harveyBallFull} & \multicolumn{1}{ c }{$-$} & \multicolumn{1}{ c }{$-$} & \multicolumn{1}{ c }{\textcolor{OliveGreen}{$\mathbf{+242\%}$}} & \multicolumn{1}{ c }{\textcolor{OliveGreen}{$\mathbf{+74\%}$}} & \multicolumn{1}{ c }{\textcolor{Maroon}{$\mathbf{-42\%}$}} & \multicolumn{1}{ c }{\textcolor{Maroon}{$\mathbf{-63\%}$}} & \multicolumn{1}{ c }{\textcolor{Maroon}{$\mathbf{-88\%}$}} & \multicolumn{1}{ c }{\textcolor{Maroon}{$\mathbf{-97\%}$}}\\ \cline{3-12}
		
		\multicolumn{1}{ c| }{$\kappa$} & \multicolumn{1}{ c }{$2.5$} & \multicolumn{1}{ |c }{$\tau_{t}^*$} & \multicolumn{1}{ c }{\harveyBallThreeQuarter} & \multicolumn{1}{ c }{$-$} & \multicolumn{1}{ c }{$-$} & \multicolumn{1}{ c }{\textcolor{Maroon}{$\mathbf{-11\%}$}} & \multicolumn{1}{ c }{$-5\%$} & \multicolumn{1}{ c }{$+5\%$} & \multicolumn{1}{ c }{$+9\%$} & \multicolumn{1}{ c }{\textcolor{OliveGreen}{$\mathbf{+17\%}$}} & \multicolumn{1}{ c }{\textcolor{OliveGreen}{$\mathbf{+32\%}$}}\\ \cline{3-12}
		
		&  & \multicolumn{1}{ |c }{$\Lambda^*$} & \multicolumn{1}{ c }{\harveyBallFull} & \multicolumn{1}{ c }{$-$} & \multicolumn{1}{ c }{$-$} & \multicolumn{1}{ c }{\textcolor{OliveGreen}{$\mathbf{+199\%}$}} & \multicolumn{1}{ c }{\textcolor{OliveGreen}{$\mathbf{+64\%}$}} & \multicolumn{1}{ c }{\textcolor{Maroon}{$\mathbf{-39\%}$}} & \multicolumn{1}{ c }{\textcolor{Maroon}{$\mathbf{-59\%}$}} & \multicolumn{1}{ c }{\textcolor{Maroon}{$\mathbf{-86\%}$}} & \multicolumn{1}{ c }{\textcolor{Maroon}{$\mathbf{-96\%}$}}\\ \cline{1-12} 
		\cline{1-12}

		&  & \multicolumn{1}{ |c }{$r_{s}^*$} & \multicolumn{1}{ c }{\harveyBallHalf} & \multicolumn{1}{ c }{\textcolor{Maroon}{$\mathbf{-14\%}$}} & \multicolumn{1}{ c }{$-7\%$} & \multicolumn{1}{ c }{$-3\%$} & \multicolumn{1}{ c }{$-1\%$} & \multicolumn{1}{ c }{$+1\%$} & \multicolumn{1}{ c }{$+3\%$} & \multicolumn{1}{ c }{$+7\%$} & \multicolumn{1}{ c }{\textcolor{OliveGreen}{$\mathbf{+15\%}$}}\\ \cline{3-12}
		
		\multicolumn{1}{ c| }{$\beta$} & \multicolumn{1}{ c }{$3.0$} & \multicolumn{1}{ |c }{$\tau_{t}^*$} & \multicolumn{1}{ c }{\harveyBallQuarter} & \multicolumn{1}{ c }{$+9\%$} & \multicolumn{1}{ c }{$+5\%$} & \multicolumn{1}{ c }{$+2\%$} & \multicolumn{1}{ c }{$+1\%$} & \multicolumn{1}{ c }{$-1\%$} & \multicolumn{1}{ c }{$-2\%$} & \multicolumn{1}{ c }{$-5\%$} & \multicolumn{1}{ c }{\textcolor{Maroon}{$\mathbf{-10\%}$}}\\ \cline{3-12}
		
		&  & \multicolumn{1}{ |c }{$\Lambda^*$} & \multicolumn{1}{ c }{\harveyBallFull} & \multicolumn{1}{ c }{\textcolor{OliveGreen}{$\mathbf{+59\%}$}} & \multicolumn{1}{ c }{\textcolor{OliveGreen}{$\mathbf{+25\%}$}} & \multicolumn{1}{ c }{$+9\%$} & \multicolumn{1}{ c }{$+4\%$} & \multicolumn{1}{ c }{$-4\%$} & \multicolumn{1}{ c }{$-8\%$} & \multicolumn{1}{ c }{\textcolor{Maroon}{$\mathbf{-19\%}$}} & \multicolumn{1}{ c }{\textcolor{Maroon}{$\mathbf{-34\%}$}}\\ \cline{1-12}

		&  & \multicolumn{1}{ |c }{$r_{s}^*$} & \multicolumn{1}{ c }{\harveyBallQuarter} & \multicolumn{1}{ c }{$+9\%$} & \multicolumn{1}{ c }{$+3\%$} & \multicolumn{1}{ c }{$+1\%$} & \multicolumn{1}{ c }{$+1\%$} & \multicolumn{1}{ c }{$-$} & \multicolumn{1}{ c }{$-1\%$} & \multicolumn{1}{ c }{$-2\%$} & \multicolumn{1}{ c }{$-3\%$}\\ \cline{3-12}
		
		\multicolumn{1}{ c| }{$\alpha$} & \multicolumn{1}{ c }{$3.0$} & \multicolumn{1}{ |c }{$\tau_{t}^*$} & \multicolumn{1}{ c }{\harveyBallThreeQuarter} & \multicolumn{1}{ c }{\textcolor{OliveGreen}{$\mathbf{+25\%}$}} & \multicolumn{1}{ c }{$+9\%$} & \multicolumn{1}{ c }{$+3\%$} & \multicolumn{1}{ c }{$+1\%$} & \multicolumn{1}{ c }{$-1\%$} & \multicolumn{1}{ c }{$-3\%$} & \multicolumn{1}{ c }{$-6\%$} & \multicolumn{1}{ c }{\textcolor{Maroon}{$\mathbf{-11\%}$}}\\ \cline{3-12}
		
		&  & \multicolumn{1}{ |c }{$\Lambda^*$} & \multicolumn{1}{ c }{\harveyBallFull} & \multicolumn{1}{ c }{\textcolor{Maroon}{$\mathbf{-43\%}$}} & \multicolumn{1}{ c }{\textcolor{Maroon}{$\mathbf{-19\%}$}} & \multicolumn{1}{ c }{$-7\%$} & \multicolumn{1}{ c }{$-3\%$} & \multicolumn{1}{ c }{$+3\%$} & \multicolumn{1}{ c }{$+6\%$} & \multicolumn{1}{ c }{\textcolor{OliveGreen}{$\mathbf{+14\%}$}} & \multicolumn{1}{ c }{\textcolor{OliveGreen}{$\mathbf{+25\%}$}}\\ \cline{1-12}

		&  & \multicolumn{1}{ |c }{$r_{s}^*$} & \multicolumn{1}{ c }{\harveyBallQuarter} & \multicolumn{1}{ c }{$-5\%$} & \multicolumn{1}{ c }{$-5\%$} & \multicolumn{1}{ c }{$-2\%$} & \multicolumn{1}{ c }{$-1\%$} & \multicolumn{1}{ c }{$+1\%$} & \multicolumn{1}{ c }{$+2\%$} & \multicolumn{1}{ c }{$+4\%$} & \multicolumn{1}{ c }{$+8\%$}\\ \cline{3-12}
		
		\multicolumn{1}{ c| }{$\rho_p$} & \multicolumn{1}{ c }{$0.001$} & \multicolumn{1}{ |c }{$\tau_{t}^*$} & \multicolumn{1}{ c }{\harveyBallHalf} & \multicolumn{1}{ c }{\textcolor{OliveGreen}{$\mathbf{+15\%}$}} & \multicolumn{1}{ c }{\textcolor{OliveGreen}{$\mathbf{+14\%}$}} & \multicolumn{1}{ c }{$+5\%$} & \multicolumn{1}{ c }{$+2\%$} & \multicolumn{1}{ c }{$-2\%$} & \multicolumn{1}{ c }{$-5\%$} & \multicolumn{1}{ c }{\textcolor{Maroon}{$\mathbf{-10\%}$}} & \multicolumn{1}{ c }{\textcolor{Maroon}{$\mathbf{-19\%}$}}\\ \cline{3-12}
		
		&  & \multicolumn{1}{ |c }{$\Lambda^*$} & \multicolumn{1}{ c }{\harveyBallQuarter} & \multicolumn{1}{ c }{\textcolor{OliveGreen}{$\mathbf{+13\%}$}} & \multicolumn{1}{ c }{$+6\%$} & \multicolumn{1}{ c }{$+2\%$} & \multicolumn{1}{ c }{$+1\%$} & \multicolumn{1}{ c }{$-1\%$} & \multicolumn{1}{ c }{$-2\%$} & \multicolumn{1}{ c }{$-5\%$} & \multicolumn{1}{ c }{$-9\%$}\\ \cline{1-12}

		&  & \multicolumn{1}{ |c }{$r_{s}^*$} & \multicolumn{1}{ c }{\harveyBallHalf} & \multicolumn{1}{ c }{\textcolor{Maroon}{$\mathbf{-18\%}$}} & \multicolumn{1}{ c }{$-9\%$} & \multicolumn{1}{ c }{$-3\%$} & \multicolumn{1}{ c }{$-2\%$} & \multicolumn{1}{ c }{$+2\%$} & \multicolumn{1}{ c }{$+4\%$} & \multicolumn{1}{ c }{$+9\%$} & \multicolumn{1}{ c }{\textcolor{OliveGreen}{$\mathbf{+19\%}$}}\\ \cline{3-12}
		
		\multicolumn{1}{ c| }{$r_p$} & \multicolumn{1}{ c }{$100$} & \multicolumn{1}{ |c }{$\tau_{t}^*$} & \multicolumn{1}{ c }{\harveyBallFull} & \multicolumn{1}{ c }{\textcolor{OliveGreen}{$\mathbf{+64\%}$}} & \multicolumn{1}{ c }{\textcolor{OliveGreen}{$\mathbf{+27\%}$}} & \multicolumn{1}{ c }{\textcolor{OliveGreen}{$\mathbf{+10\%}$}} & \multicolumn{1}{ c }{$+5\%$} & \multicolumn{1}{ c }{$-5\%$} & \multicolumn{1}{ c }{$-9\%$} & \multicolumn{1}{ c }{\textcolor{Maroon}{$\mathbf{-20\%}$}} & \multicolumn{1}{ c }{\textcolor{Maroon}{$\mathbf{-36\%}$}}\\ \cline{3-12}
		
		&  & \multicolumn{1}{ |c }{$\Lambda^*$} & \multicolumn{1}{ c }{\harveyBallHalf} & \multicolumn{1}{ c }{\textcolor{OliveGreen}{$\mathbf{+23\%}$}} & \multicolumn{1}{ c }{\textcolor{OliveGreen}{$\mathbf{+11\%}$}} & \multicolumn{1}{ c }{$+4\%$} & \multicolumn{1}{ c }{$+2\%$} & \multicolumn{1}{ c }{$-2\%$} & \multicolumn{1}{ c }{$-4\%$} & \multicolumn{1}{ c }{\textcolor{Maroon}{$\mathbf{-10\%}$}} & \multicolumn{1}{ c }{\textcolor{Maroon}{$\mathbf{-19\%}$}}\\ \cline{1-12}

		&  & \multicolumn{1}{ |c }{$r_{s}^*$} & \multicolumn{1}{ c }{\harveyBallThreeQuarter} & \multicolumn{1}{ c }{\textcolor{OliveGreen}{$\mathbf{+32\%}$}} & \multicolumn{1}{ c }{\textcolor{OliveGreen}{$\mathbf{+12\%}$}} & \multicolumn{1}{ c }{$+4\%$} & \multicolumn{1}{ c }{$+2\%$} & \multicolumn{1}{ c }{$-2\%$} & \multicolumn{1}{ c }{$-4\%$} & \multicolumn{1}{ c }{$-8\%$} & \multicolumn{1}{ c }{\textcolor{Maroon}{$\mathbf{-12\%}$}}\\ \cline{3-12}
		
		\multicolumn{1}{ c| }{$\gamma_0$} & \multicolumn{1}{ c }{$20$ dB} & \multicolumn{1}{ |c }{$\tau_{t}^*$} & \multicolumn{1}{ c }{\harveyBallNone} & \multicolumn{1}{ c }{$-$} & \multicolumn{1}{ c }{$-$} & \multicolumn{1}{ c }{$-$} & \multicolumn{1}{ c }{$-$} & \multicolumn{1}{ c }{$-$} & \multicolumn{1}{ c }{$-$} & \multicolumn{1}{ c }{$-$} & \multicolumn{1}{ c }{$-$}\\ \cline{3-12}
		
		&  & \multicolumn{1}{ |c }{$\Lambda^*$} & \multicolumn{1}{ c }{\harveyBallQuarter} & \multicolumn{1}{ c }{\textcolor{OliveGreen}{$\mathbf{+12\%}$}} & \multicolumn{1}{ c }{$+5\%$} & \multicolumn{1}{ c }{$+2\%$} & \multicolumn{1}{ c }{$+1\%$} & \multicolumn{1}{ c }{$-1\%$} & \multicolumn{1}{ c }{$-2\%$} & \multicolumn{1}{ c }{$-4\%$} & \multicolumn{1}{ c }{$-7\%$}\\ \cline{1-12}

		&  & \multicolumn{1}{ |c }{$r_{s}^*$} & \multicolumn{1}{ c }{\harveyBallHalf} & \multicolumn{1}{ c }{\textcolor{Maroon}{$\mathbf{-24\%}$}} & \multicolumn{1}{ c }{\textcolor{Maroon}{$\mathbf{-11\%}$}} & \multicolumn{1}{ c }{$-4\%$} & \multicolumn{1}{ c }{$-2\%$} & \multicolumn{1}{ c }{$+2\%$} & \multicolumn{1}{ c }{$+4\%$} & \multicolumn{1}{ c }{$+9\%$} & \multicolumn{1}{ c }{\textcolor{OliveGreen}{$\mathbf{+18\%}$}}\\ \cline{3-12}
		
		\multicolumn{1}{ c| }{$P_s$} & \multicolumn{1}{ c }{$700$ mW} & \multicolumn{1}{ |c }{$\tau_{t}^*$} & \multicolumn{1}{ c }{\harveyBallNone} & \multicolumn{1}{ c }{$-1\%$} & \multicolumn{1}{ c }{$-$} & \multicolumn{1}{ c }{$-$} & \multicolumn{1}{ c }{$-$} & \multicolumn{1}{ c }{$-$} & \multicolumn{1}{ c }{$-$} & \multicolumn{1}{ c }{$-$} & \multicolumn{1}{ c }{$-$}\\ \cline{3-12}
		
		&  & \multicolumn{1}{ |c }{$\Lambda^*$} & \multicolumn{1}{ c }{\harveyBallFull} & \multicolumn{1}{ c }{\textcolor{OliveGreen}{$\mathbf{+51\%}$}} & \multicolumn{1}{ c }{\textcolor{OliveGreen}{$\mathbf{+19\%}$}} & \multicolumn{1}{ c }{$+7\%$} & \multicolumn{1}{ c }{$+3\%$} & \multicolumn{1}{ c }{$-3\%$} & \multicolumn{1}{ c }{$-6\%$} & \multicolumn{1}{ c }{\textcolor{Maroon}{$\mathbf{-13\%}$}} & \multicolumn{1}{ c }{\textcolor{Maroon}{$\mathbf{-22\%}$}}\\ \cline{1-12}

		&  & \multicolumn{1}{ |c }{$r_{s}^*$} & \multicolumn{1}{ c }{\harveyBallQuarter} & \multicolumn{1}{ c }{$+7\%$} & \multicolumn{1}{ c }{$+3\%$} & \multicolumn{1}{ c }{$+1\%$} & \multicolumn{1}{ c }{$-$} & \multicolumn{1}{ c }{$-$} & \multicolumn{1}{ c }{$-1\%$} & \multicolumn{1}{ c }{$-2\%$} & \multicolumn{1}{ c }{$-4\%$}\\ \cline{3-12}
		
		\multicolumn{1}{ c| }{$P_{col}$} & \multicolumn{1}{ c }{$0.04$} & \multicolumn{1}{ |c }{$\tau_{t}^*$} & \multicolumn{1}{ c }{\harveyBallHalf} & \multicolumn{1}{ c }{\textcolor{OliveGreen}{$\mathbf{+13\%}$}} & \multicolumn{1}{ c }{$+6\%$} & \multicolumn{1}{ c }{$+2\%$} & \multicolumn{1}{ c }{$+1\%$} & \multicolumn{1}{ c }{$-1\%$} & \multicolumn{1}{ c }{$-2\%$} & \multicolumn{1}{ c }{$-5\%$} & \multicolumn{1}{ c }{$-9\%$}\\ \cline{3-12}
		
		&  & \multicolumn{1}{ |c }{$\Lambda^*$} & \multicolumn{1}{ c }{\harveyBallThreeQuarter} & \multicolumn{1}{ c }{\textcolor{Maroon}{$\mathbf{-30\%}$}} & \multicolumn{1}{ c }{\textcolor{Maroon}{$\mathbf{-14\%}$}} & \multicolumn{1}{ c }{$-6\%$} & \multicolumn{1}{ c }{$-3\%$} & \multicolumn{1}{ c }{$+3\%$} & \multicolumn{1}{ c }{$+6\%$} & \multicolumn{1}{ c }{\textcolor{OliveGreen}{$\mathbf{+14\%}$}} & \multicolumn{1}{ c }{\textcolor{OliveGreen}{$\mathbf{+27\%}$}}\\ \cline{1-12}

		&  & \multicolumn{1}{ |c }{$r_{s}^*$} & \multicolumn{1}{ c }{\harveyBallQuarter} & \multicolumn{1}{ c }{$+7\%$} & \multicolumn{1}{ c }{$+2\%$} & \multicolumn{1}{ c }{$+1\%$} & \multicolumn{1}{ c }{$-$} & \multicolumn{1}{ c }{$-$} & \multicolumn{1}{ c }{$-$} & \multicolumn{1}{ c }{$-1\%$} & \multicolumn{1}{ c }{$-2\%$}\\ \cline{3-12}
		
		\multicolumn{1}{ c |}{$\sigma_p^2/\sigma_n^2$} & \multicolumn{1}{ c }{$10$} & \multicolumn{1}{ |c }{$\tau_{t}^*$} & \multicolumn{1}{ c }{\harveyBallHalf} & \multicolumn{1}{ c }{\textcolor{OliveGreen}{$\mathbf{+16\%}$}} & \multicolumn{1}{ c }{$+6\%$} & \multicolumn{1}{ c }{$+2\%$} & \multicolumn{1}{ c }{$+1\%$} & \multicolumn{1}{ c }{$-1\%$} & \multicolumn{1}{ c }{$-2\%$} & \multicolumn{1}{ c }{$-4\%$} & \multicolumn{1}{ c }{$-6\%$}\\ \cline{3-12}
		
		&  & \multicolumn{1}{ |c }{$\Lambda^*$} & \multicolumn{1}{ c }{\harveyBallThreeQuarter} & \multicolumn{1}{ c }{\textcolor{Maroon}{$\mathbf{+32\%}$}} & \multicolumn{1}{ c }{\textcolor{Maroon}{$\mathbf{-12\%}$}} & \multicolumn{1}{ c }{$-4\%$} & \multicolumn{1}{ c }{$-2\%$} & \multicolumn{1}{ c }{$+2\%$} & \multicolumn{1}{ c }{$+4\%$} & \multicolumn{1}{ c }{$+9\%$} & \multicolumn{1}{ c }{\textcolor{OliveGreen}{$\mathbf{+15\%}$}}\\ \cline{1-12}
		
		\multicolumn{12}{ c }{\textbf{Legend} \harveyBallNone: Not sensitive at all $\|$ \harveyBallQuarter: Slightly sensitive $\|$ \harveyBallHalf: Mildly sensitive $\|$ \harveyBallThreeQuarter: Significantly sensitive $\|$ \harveyBallFull: Very sensitive} 	\\	\cline{1-12}
	\end{tabularx}
} 
	\egroup
	\label{table:Sensitivity}
\end{table*} 

\subsubsection{Sensitivity to $\kappa$}
\label{sec:Kappa}
Path loss exponent $\kappa$ is the parameter that optimal ($r_{s}^*$, $\tau_{t}^*$, $\Lambda^*$) are most sensitive to. When $\kappa$ is raised, all else being equal (incl. base value ($r_{s}^*$, $\tau_{t}^*$)), $E\{E_{sft}\}$ increases and achievable rate $R$ decreases, as shown in Table \ref{table:Sensitivity sidetable}. Increase in $E\{E_{sft}\}$ is very significant, driven by the fact that consumed energy in transmission is $E_t=(q_1 r_{s}^{\kappa} + q_2)\tau_t$. To that end, it is clear that new optimal $\Lambda^*$ that to be attained once ($r_{s}^*$, $\tau_{t}^*$) are updated will be much lower.

Given the $E_t$ equation form, to arrive at the new optimal $\Lambda^*$, shift in $r_{s}^*$ in the reverse direction of the change in $\kappa$ is more effective compared to a change in $\tau_{t}^*$. Hence, $r_{s}^*$ moves significantly in the reverse direction of the $\kappa$ change. Albeit not at the order of $r_{s}^*$, $\tau_{t}^*$ also shifts significantly with changes in $\kappa$. However, in contrast to $r_{s}^*$, $\tau_{t}^*$ moves in the same direction as $\kappa$, as indicated in rows 1-3 of Table \ref{table:Sensitivity}.

\subsubsection{Sensitivity to $\beta$}
\label{sec:Beta}
PU birth rate $\beta$ is the second parameter that optimal $\Lambda^*$ is most sensitive to. All else being equal (incl. base value ($r_{s}^*$, $\tau_{t}^*$)), a rise in $\beta$ increases both $E\{\tau_{sft}\}$ and $E\{E_{sft}\}$, as given in Table \ref{table:Sensitivity sidetable}. This is driven by the fact that a surge in $\beta$ drives down both $P_{idle}$ probability for SSA as well as $P_{np}$ probability for SFT once SSA is achieved.

Adjustments on ($r_{s}$, $\tau_{t}$) have little to no impact on $P_{idle}$ and $P_{fa}$ probabilities that drive the likelihood of SSA. However, both decision variables are directly relevant for $P_{np}=e^{-\rho_{p}S (1-e^{-\tau_{t} \beta})}$, which is the relevant probability to complete SFT after SSA is achieved. To optimize the reward function, system chooses to increase $E\{W\}$ via increasing $r_{s}$ and reduce chances for PU arrival during frame transmission by decreasing $\tau_{t}$, as indicated in rows 4-6 of Table \ref{table:Sensitivity}. Consequently, both $r_{s}^*$ and $\tau_{t}^*$ shift respectably as a response to changes in $\beta$. $r_{s}^*$ moves in the same direction as $\beta$, while $\tau_{t}^*$ shifts to the reverse. It is worth noting that an increase in $r_{s}^*$ is possible here, despite an increase in $\beta$, driven by the low base value for PU node density $\rho_p$.

\subsubsection{Sensitivity to $\alpha$}
\label{sec:Alpha}
PU death rate $\alpha$ is the third parameter that optimal reward is most sensitive to. When $\alpha$ rises, both $E\{\tau_{sft}\}$ and $E\{E_{sft}\}$ decrease, as revealed in Table \ref{table:Sensitivity sidetable}. This is driven by the fact that a surge in $\alpha$ boosts $P_{idle}$ hence likelihood of SSA.

To arrive at new optimal $\Lambda^*$, since average idle time of each channel is distributed exponentially with mean $1/\beta$, there is no reason for $\tau_{t}$ to increase as $\alpha$ increases. On the other hand, an increase in $\alpha$ surges $P_{idle}$ to help SSA, yet has no impact on $P_{np}$ that is relevant for SFT after SSA is achieved. Therefore, optimization seeks to maximize $P_{np}=e^{-\rho_{p}S (1-e^{-\tau_{t} \beta})}$, which is the probability that no PU arrives to the channel during frame transmission. It is much easier to raise this probability by decreasing $\tau_{t}$ rather than $r_{s}$, as $S$ is a slowly increasing function with respect to $r_s$. To that end, all else being equal, $\tau_{t}^*$ and $r_{s}^*$ are negatively correlated with $\alpha$, with $\tau_{t}^*$ having a much higher sensitivity, as in rows 7-9 of Table \ref{table:Sensitivity}.

\subsubsection{Sensitivity to $\rho_p$}
\label{sec:Rho_p} 
Optimal ($r_{s}^*$, $\tau_{t}^*$, $\Lambda^*$) are slightly sensitive to changes in PU node density $\rho_p$. All else being equal (incl. base value ($r_{s}^*$, $\tau_{t}^*$)) when $\rho_p$ is increased, both $E\{\tau_{sft}\}$ and $E\{E_{sft}\}$ rise, as tabulated in Table \ref{table:Sensitivity sidetable}. This in turn decreases the attainable $\Lambda$ value.

Parameter $\rho_p$ essentially has no direct link to likelihood of SSA, yet is impacting $P_{np}=e^{-\rho_{p}S (1-e^{-\tau_{t} \beta})}$ and hence SFT probability through the $S$ term, which is defined as $S(w,r_p)=2\pi r_p^2 - 2r_p cos^{-1}\big(\frac{w}{2r_p}\big) + \frac{w}{2}\sqrt{4r_p^2-w^2}$. Here, $E\{W\}=w$ for a given $r_s$. As shown in rows 10-12 of Table \ref{table:Sensitivity}, when $r_p$ is changed, to arrive at new optimal $\Lambda^*$, $\tau_{t}^*$ is mildly shifted in the opposite direction while $r_{s}^*$ is slightly moved in the same direction. In that sense, as $\rho_p$ rises, system chooses progressing more meters toward the sink at each hop at the expense of reduced transmission time. It should be noted that an increase in $r_{s}^*$ is possible here, despite an increase in $\rho_p$, driven by the low base value for $\rho_p$.

\subsubsection{Sensitivity to $r_p$}
\label{sec:r_p} 
Optimal ($r_{s}^*$, $\tau_{t}^*$, $\Lambda^*$) are slightly sensitive to changes in PU guardring radius $r_p$ . All else being equal (incl. base value ($r_{s}^*$, $\tau_{t}^*$)) when $r_p$ is increased, both $E\{\tau_{sft}\}$ and $E\{E_{sft}\}$ rise, as in Table \ref{table:Sensitivity sidetable}. It is clear that reward value decreases with increasing $r_p$.

$r_p$ has a similar story as $\rho_p$, having no direct link to likelihood of SSA but impacting the $P_{np}$ probability for SFT completion. Hence, as a response to changes in $r_p$, $\tau_{t}^*$ is significantly shifted in the opposite direction while $r_{s}^*$ is mildly moved in the same direction, as disclosed in rows 13-15 of Table \ref{table:Sensitivity}. In fact, $\tau_{t}^*$ is most sensitive to $r_p$ out of all the 9 parameters in scrutiny.

\subsubsection{Sensitivity to $\gamma_0$}
\label{sec:gamma_0} 
Optimal $\Lambda^*$ is slightly and $r_{s}^*$ is significantly sensitive to changes in $\gamma_0$, while $\tau_{t}^*$ is not impacted at all. When $\gamma_0$ is increased, all else being equal (incl. base value ($r_{s}^*$, $\tau_{t}^*$)), both the achievable rate $R$ and $E\{E_{sft}\}$ increase, as seen in Table \ref{table:Sensitivity sidetable}.

$\gamma_0$ has no direct link to SSA or SFT probabilities, yet mainly impacts the consumed energy during attempted transmissions, $E_t=(q_1 r_{s}^{\kappa} + q_2)\tau_t$. $\gamma_0$ is a linear multiplier term within $q_1$. As $\gamma_0$ is changed, to arrive at the new optimal $\Lambda^*$, system prefers to adjust $r_{s}^*$ in the opposite direction of the change of $\gamma_0$, rather than $\tau_{t}^*$. This is driven by the fact that $r_s$ has an exponent $\kappa=2.5$ in the $E_t$ equation, whereas $\tau_t$ is a linear multiplier. Hence, $r_{s}^*$ is significantly sensitive to and negatively correlated with $\gamma_0$, while $\tau_{t}^*$ is uncorrelated, as shown in rows 16-18 of Table \ref{table:Sensitivity}.

\subsubsection{Sensitivity to $P_s$}
\label{sec:P_S}
Optimal $\Lambda^*$ is very sensitive to power consumption during spectrum sensing $P_s$. When $P_s$ is increased, all else being equal, only $E\{E_{sft}\}$ increases, while no impact is observed on other terms of the $\Lambda^*$ expression from (\ref{eq:Reward function}), as seen in Table \ref{table:Sensitivity sidetable}.

All three possible scenario sets $A$, $B$ and $C$ contain the term $2E_s = 2P_s\tau_s$ in their energy consumption in each slot. To that end, $2E_s$ is sort of a fixed cost in $E\{E_{sft}\}$ which is consumed until SFT. To that end, when $P_s$ is increased, the system seeks to arrive at the new optimal $\Lambda^*$ via increasing $r_{s}^*$. This increase in $r_{s}^*$ is driven by the fact that system would like to achieve more of either $\tau_t R$ or $E\{W\}$ as fixed cost in the energy component $E\{E_{sft}\}$ becomes even more important. Raising $\tau_t R$ via increasing $\tau_{t}^*$ is not very feasible, as this increases the likelihood of failed transmissions without SFT, thus increasing $E\{\tau_{sft}\}$ and $E\{E_{sft}\}$. In contrast, $r_{s}^*$ is linked to SFT probability through the $S$ term in $P_{np}=e^{-\rho_{p}S (1-e^{-\tau_{t} \beta})}$, which is a slowly increasing function of $r_{s}$. Hence, $r_{s}^*$ shifts significantly in the same direction as $\gamma_0$, whereas $\tau_{t}^*$ remains intact, as shown in rows 19-21 of Table \ref{table:Sensitivity}.

\subsubsection{Sensitivity to $P_{col}$}
\label{sec:P_col}
Optimal $\Lambda^*$ is significantly sensitive to the maximum SU-PU collision probability that a PU can tolerate $P_{col}$.  When $P_{col}$ is increased, the constraint on maximum SU-PU collision probability is relaxed. Consequently, all else being equal (incl. base value ($r_{s}^*$, $\tau_{t}^*$)), both $E\{\tau_{sft}\}$ and $E\{E_{sft}\}$ decrease as seen in Table \ref{table:Sensitivity sidetable}, increasing $\Lambda$ as a consequence.

We have seen in Fig. \ref{fig:Figure_5}(a) that as $P_{col}$ decreases, required $\tau_s$ significantly increases. Consequently, this increases not only the fixed cost component of energy consumption of amount $2E_s$ but also the time consumed for spectrum sensing $\tau_s$, in each slot. To that end, as $P_{col}$ decreases, to arrive at the new optimal $\Lambda^*$, the system shifts both $r_{s}^*$ and $\tau_{t}^*$ in the reverse direction. $r_{s}^*$ is slightly and $\tau_{t}^*$ is mildly sensitive to changes in $P_{col}$, as shown in rows 22-24 of Table \ref{table:Sensitivity}.

\subsubsection{Sensitivity to $\sigma_p^2/\sigma_n^2$}
\label{sec:Sigma}
Optimal $\Lambda^*$ is significantly sensitive to PU signal variance as a ratio to noise variance $\sigma_{p}^2/\sigma_{n}^2$. When $\sigma_{p}^2/\sigma_{n}^2$ is increased, all else being equal (incl. base value ($r_{s}^*$, $\tau_{t}^*$)), both $E\{\tau_{sft}\}$ and $E\{E_{sft}\}$ decrease as shown in Table \ref{table:Sensitivity sidetable}, raising $\Lambda$ as a result.

Assuming $\sigma_{n}^2$ fixed, increasing $\sigma_{p}^2$ for a given $\tau_s$ decreases the false alarm probability $P_{fa}$ in (\ref{eq:P_fa}) and the mis-detection probability $P_{md}$ in (\ref{eq:P_md}). Hence, to achieve a fixed target $P_{fa}=P_{md}$, $\tau_s$ can be reduced when $\sigma_{p}^2/\sigma_{n}^2$ is increased. This curtails not only the $2E_s$ amount of energy consumption but also the time for spectrum sensing $\tau_s$, in each slot. Hence, $r_{s}^*$ and $\tau_{t}^*$ are negatively correlated with $\sigma_{p}^2/\sigma_{n}^2$, with $\tau_{t}^*$ being more sensitive (in rows 25-27 of Table \ref{table:Sensitivity}).

\subsection{Sensitivity of $r_{s}^*$, $\tau_{t}^*$ and reward to $\kappa$, $\beta$ and $\alpha$}
\label{sec:Sensitivity with K and alpha}
\begin{figure}[!t]
	\centering
	\includegraphics[width=0.99\columnwidth]{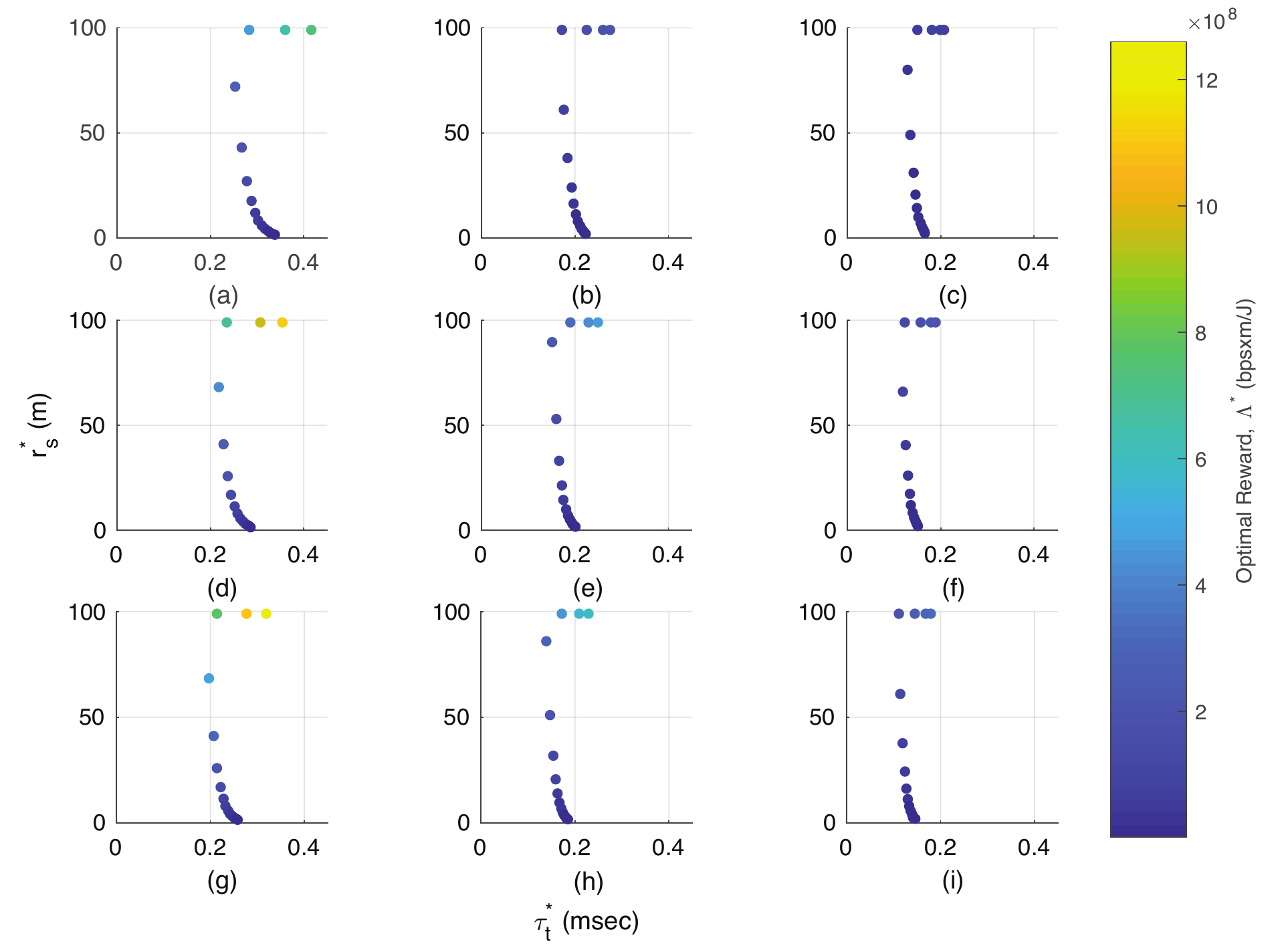}
	\caption{ Optimal ($r_{s}^*$, $\tau_{t}^*$, $\Lambda^*$) with respect to (a) $\alpha=10$, $\beta=10$; (b) $\alpha=10$, $\beta=20$; (c) $\alpha=10$, $\beta=30$; (d) $\alpha=20$, $\beta=10$; (e) $\alpha=20$, $\beta=20$; (f) $\alpha=20$, $\beta=30$; (g) $\alpha=30$, $\beta=10$; (h) $\alpha=30$, $\beta=20$; (i) $\alpha=30$, $\beta=30$, when $\kappa$ changes from $2.0$ to $3.4$ with $0.1$ steps, and PU guardring radius set as $r_p=100$ m,}
	\label{fig:Figure_8}
\end{figure}

As revealed in Table \ref{table:Sensitivity}, $\Lambda$ is most sensitive to changes in $\kappa$, $\beta$ and $\alpha$. Hence, we investigate the combined effect of changes in $\kappa$, $\beta$ and $\alpha$ to ($r_{s}^*$, $\tau_{t}^*$, $\Lambda^*$) in Fig. \ref{fig:Figure_8}. In this representation, each row stands for a fixed $\alpha$ value ($\alpha=10$, $\alpha=20$ and $\alpha=30$ for first, second and third rows, respectively) and each column stands for a fixed $\beta$ value ($\beta=10$, $\beta=20$ and $\beta=30$ for first, second and third columns, respectively). Additionally, each sub-figure displays 15 data points, each corresponding to a case on $\kappa=2$ to $\kappa=3.4$ with $0.1$ steps. We also fix the PU guardring radius as $r_p=100$m, with the condition that $r_s \leq r_p$, which is the usual case for CRSNs.

Accordingly, each sub-figure starts the data points with $\kappa=2.0$ scenario, which yields $r_{s}^*=100$ m in all 9 cases. Again in all cases, $r_{s}^*$ stays at $100$ m for 3-4 data points corresponding to $\kappa=2.0, 2.1, 2.2, 2.3$ scenarios, while $\tau_{t}^*$ is decreased. After $\kappa=2.4$ or $\kappa=2.5$ depending on the case, $r_{s}^*$ is consistently reduced and $\tau_{t}^*$ is consistently boosted toward $\kappa=3.4$ scenario.

As previously revealed in Section \ref{sec:Kappa}, $r_{s}^*$ moves in a very sensitive manner in the reverse direction of the change in $\kappa$, whereas $\tau_{t}^*$ moves significantly in the same direction as $\kappa$. Hence, in each sub-figure, we see an increase in $r_{s}^*$ until $100$ m, as $\kappa$ is decreased from $3.4$ to $2.3$-$2.4$. After that point, further decrease in $\kappa$ cannot trigger any more $r_{s}^*$ increase as it is capped by $r_p=100$ m. Therefore, we observe an increase in $\tau_{t}^*$ as we further reduce $\kappa$ from $2.3$-$2.4$ levels to $2.0$, despite they are being positively correlated under normal circumstances. Key highlights on sensitivities of ($r_{s}^*$, $\tau_{t}^*$, $\Lambda^*$) on $\kappa$, $\beta$ and $\alpha$ are as follows:
\begin{itemize}
\item All three are most sensitive to $\kappa$. $r_{s}^*$ and $\Lambda^*$ move in reverse direction of $\kappa$, whereas $\tau_{t}^*$ follows $\kappa$ with the exception of when $r_{s}^*$ is capped to $100$ m.
\item Increasing $\beta$ decreases $\tau_{t}^*$ significantly irrespective of $\kappa$ value (except when $r_{s}^*$ is capped to $100$ m). Assuming $\alpha=10$, increasing $\beta$ from $10$ to $20$ and from $10$ t $30$ reduces $\tau_{t}^*$ by $\sim 33\%$ and $\sim 50\%$, respectively. Boosting $\alpha$ smoothens this decrease. As an example, when $\alpha=30$, $\tau_{t}^*$ decreases by $\sim 29\%$ and $\sim 45\%$ for the same increases in $\beta$.
\item Increasing $\alpha$ decreases $\tau_{t}^*$ significantly irrespective of $\kappa$ value. If $\beta=10$, increasing $\alpha$ from $10$ to $20$ and from $10$ to $30$ reduces $\tau_{t}^*$ by $15\%$ and $24\%$, respectively. Boosting $\beta$ smoothens this decrease. When $\beta=30$, $\tau_{t}^*$ decreases by $9\%$ and $13\%$ for the same increases in $\alpha$.
\end{itemize}

\section{Conclusions}
In this paper, we formulate the problem of finding energy-efficient transmission range $r_s$ and transmission duration $\tau_t$ for CRSN deployment that would minimize the energy consumed per goodput per meter toward the sink in a greedy forwarding scenario. For that purpose, we first characterize the transmissions in CRSN under DSA; deriving the probability of successful frame transmission at each attempt between one-hop neighbor CRSN nodes. Leveraging expected hop progress toward the sink and expected hop distance concepts, we reveal the expected time and energy consumption until successful frame transmission under different spectrum utilization scenario probabilities. Finally, we formulate our optimization problem as a reward function $\Lambda$ that represents goodput times expected hop progress toward the sink per Joule. Our findings reveal that
\begin{itemize}
\item Path loss exponent $\kappa$ is by far the parameter that $r_{s}^*$ and $\Lambda^*$ are most sensitive to,
\item Apart from $\kappa$, $\Lambda^*$ is also very sensitive to PU birth rate $\beta$, PU death rate $\alpha$, SU power consumption during spectrum sensing $P_s$, maximum SU-PU collision probability that a PU can tolerate $P_{col}$ and PU signal variance $\sigma_{p}^2$,
\item Apart from $\kappa$, $r_{s}^*$ is also very sensitive to reference SNR needed at SU receiver for demodulation $\gamma_0$ and SU power consumption during spectrum sensing $P_s$,
\item $\tau_{t}^*$ is impacted by $\kappa$, PU guardring radius $r_p$, PU death rate $\alpha$ and PU node density $\rho_p$.
\end{itemize}
These relations provide valuable insights for detailed CRSN design prior to deployment. {As future work, we will investigate effects of noise variance in channel sensing and fading channel on energy-efficient transmission range and duration, which require further analysis on bit error rates and algorithm designs on the spectrum sensing.}
\label{Conclusions}

\end{document}